\documentclass{IWCOMP}
\usepackage[ruled]{algorithm2e}
\newcommand{\MNOTE}[1]{}

\usepackage{graphics}
\usepackage{multirow}
\usepackage{array}
\usepackage{eurosym}
\usepackage{natbib}
\usepackage{layout}

\copyrightyear{2014}

\DOI{iwc/iwu016}

\begin{document}
%\layout{}

\title[Wizard of Oz for Language Technology Applications]{Wizard of Oz Experimentation for Language Technology Applications: Challenges and Tools}

\author{Stephan Schl\"{o}gl}
\affiliation{Department of Management, Communication \& IT, MCI Management Center Innsbruck, Innsbruck, Austria}
\author{Gavin Doherty\corrau{Gavin.Doherty@tcd.ie}}
\author{Saturnino Luz}
\shortauthors{S. Schl\"{o}gl, G. Doherty \& S. Luz}

\affiliation{School of Computer Science and Statistics, Trinity College, The University of Dublin, Ireland}

\shortauthors{Schl\"{o}gl, Doherty \&  Luz}

%\author{}
%\affiliation{}
\shortauthors{}

\begin{abstract}
Wizard of OZ (WOZ) is a well-established method for simulating the functionality and user experience of future systems. Using a human wizard to mimic certain operations of a potential system is particularly useful in situations where extensive engineering effort would otherwise be needed to explore the design possibilities offered by such operations.
The WOZ method has been widely used in connection with speech and language technologies, but advances in sensor technology and pattern recognition as well as new application areas such as human-robot interaction have made it increasingly relevant to the design of a wider range of interactive systems. In such cases achieving acceptable performance at the user interface level often hinges on resource intensive improvements such as domain tuning, which are better done once the overall design is relatively stable. 
While WOZ is recognised as a valuable prototyping technique, surprisingly little effort has been put into exploring it from a methodological point of view. Starting from a survey of the literature, this paper presents a systematic investigation and analysis of the design space for WOZ for language technology applications, and proposes a generic architecture for tool support that supports the integration of components for speech recognition and synthesis as well as for machine translation. This architecture is instantiated in WebWOZ - a new web-based open-source WOZ prototyping platform. The viability of generic support is explored empirically through a series of evaluations. Researchers from a variety of backgrounds were able to create experiments, independent of their previous
experience with WOZ. The approach was further validated through a number of real experiments, which also helped to identify a number of possibilities for additional support, and flagged potential issues relating to consistency in Wizard performance. \end{abstract}


\category{natural language processing; design tools and techniques}

\editorial{Name}

\maketitle

% =============================================================================
\section{Introduction}\label{sec1}\vspace*{5pt}
% =============================================================================
Obtaining feedback early in the design process is important for developing high quality interactive systems.~\cite{Gou85} identified `Iterative Design' as one of three key principles for developing usable products and argue that problems and design faults can be discovered and consequently fixed through early and ongoing user testing. Prototypes, either physical or in the form of software, are valuable instruments for eliciting this sort of user feedback. Examples include paper prototypes \citep{Bai08}, sketches \citep{Kie10}, and wire-frames \citep{Li10} as well as 3D prototypes \citep{Seq05} and more advanced mock-ups \citep{Ale10}. Wizard of Oz (WOZ) is an important prototyping method used by researchers and designers to obtain feedback on functionalities that would otherwise require significant resources to be implemented. In a WOZ experiment a human `wizard' mimics the functions of a system, either entirely or in part, which permits the evaluation of potential user experiences and interaction strategies without the need for building a fully functional product first \citep{Gou83}. While WOZ can be applied in a variety of interaction scenarios, ranging from mixed-reality simulations \citep{Dow05a} to human-robot interaction \citep{Sai11}, it is mainly in the area of speech and Natural Language Processing (NLP) where the method is regularly employed, and where we see an even greater demand for it in the future. 

The reason for this expected increase can be found in the fact that the use of language technologies such as Automatic Speech Recognition (ASR), Machine Translation (MT) and Text-to-Speech Synthesis (TTS) has risen significantly in recent years. One driver of adoption has been increasingly ubiquitous access to products and services outside traditional office environments, where in many cases language technology solutions offer distinct advantages (e.g. hands-free and eyes-free interaction such as the use of speech to control a mobile phone).

Another contributing factor is the improved performance of these technologies which has opened up new application areas in  different fields. This trend is visible both from an application perspective, in the widespread use of voice dialing, in-car navigation systems with speech interfaces, instant web-based machine translation from mobile devices, and transactions accessed through Interactive Voice Response (IVR) systems, as well as from a research perspective, in emerging areas such as speech-to-speech translation \citep{Stu06} and human-avatar interaction \citep{Bra10}.

However, the technology at hand is not perfect and typically substantial engineering effort (gathering of corpora, training, tuning) is needed before prototypes involving such technologies can deliver a user experience robust enough to allow potential applications to be evaluated with real users. For Graphical User Interfaces (GUI), methods like sketching and wire-framing enable the designer to obtain early impressions and initial user feedback on a given application scenario. However, these low-fidelity prototyping techniques do not map well onto systems based around speech and other forms of natural language. Applications that use Language Technology Components such as ASR, MT or TTS as their predominant interaction channel require a different design approach and WOZ can be seen as a method that offers a means of `sketching' language-based interaction.

A review of the literature supports the view that WOZ is strongly associated with the design of interfaces that include natural language components, and related tasks such as the gathering of corpora.
The ACM Guide to Computing Literature, which contains over 2.1 million bibliographic entries at the time of writing\footnote{http://dl.acm.org/ [Accessed: Dec. 23$^{rd}$ 2013]} lists 2,045 hits for the search term `Wizard of Oz', of which 35.6\% (727 hits)  also include the keyword `Natural Language', 25.4\% the keyword `Dialog(ue) System(s)' (520 hits), 25.8\% the keywords `Corpus' or `Corpora'  (528 hits), 29.3\% the keyword `Speech recognition' (600 hits) and 59.8\% the keyword `Speech' (1,222 hits). 653 entries for Wizard of Oz did not contain any of these keywords (31.9\%). The IEEE Xplore Digital Library (over 3.6 million records at time of writing)\footnote{http://ieeexplore.ieee.org/Xplore/guesthome.jsp [Accessed: Dec. 23$^{rd}$ 2013]} lists 885 entries for `Wizard of Oz', 33.6\% (297 hits) of which also contain the term `Natural Language', 35.6\% (315 hits) `Dialog(ue) System(s)', 28.4\% (251 hits) the terms `Corpus' or `Corpora', 40.3\% (357 hits) the term `Speech Recognition' and 68.1\% (603 hits) the term `Speech'.  248 entries did not contain any of these keywords (28\%). Papers including the term `Machine Translation' are less common with 50 hits in the ACM library and 31 hits in the IEEE library, with most of these being recent. While early applications of WOZ mainly focused on simulating natural language interaction based on pure text or speech, we do see a shift towards multi-modality in recent years. However, even within this shift it is the NLP aspect of a study that typically needs the most simulation, as existing technology is simply not mature enough to be used without significant upfront investment.

Within this language-centred application area we can identify three distinct uses of the WOZ technique for designing interactive systems. Firstly, within interaction design, it is clearly possible to apply this approach to investigate the design of human-computer dialogues. Secondly, it can be used as a means for collecting language corpora  (which feeds into both interaction design and engineering work to train and tune technology components), and thirdly researchers developing technology components can employ it as a means for conducting evaluations of their performance in specific application areas, without facing the engineering effort of constructing the application itself (which may require more robust components than are currently available). 

In order to expand on this general classification of WOZ prototyping for Language Technology applications this paper presents a systematic analysis of the existing design space. This analysis is based on an extensive survey of the literature, semi-structured interviews with researchers from industry and academia who are actively involved in WOZ studies, and the requirements of researchers from a large collaborative project focused on language technology development.
\MNOTE{SL: this bit in square brackets is actually missing from the paper. I think we shoud add a section after the interviews section mentioning the actual CNGL research requirements that influenced the analysis, such as Anne's experiments, Jo\~ao's work, the need to integrate MT (cite our MT journal paper) etc. This could come before section 3, and perhaps borrow some material from it. Or do you think there's a reason to omit such inputs?}%SS>
\MNOTE{SS:Is not the last paragraph of Section 4 takling about that? I've extended it to include the references. Not sure if we need a separate section.}
%We are especially interested in better understanding the task of the wizard and how it can be supported.
 After an initial overview of possible application scenarios we move on to describing the two categories of software programs that currently support the WOZ method with respect to language technologies. Looking at the wizard task and its interplay with technology components we then analyse possible improvements in terms of tool support. In the second half of the paper we present WebWOZ - a web-based open-source WOZ prototyping platform, and empirically explore the viability of generic support through a series of evaluations involving both the construction and execution of experiments.  

It should be noted that even though the following analysis focuses mainly on WOZ prototyping for language technology applications, we believe that most, if not all, of the identified aspects generalise to other technologies and related studies, and therefore should be seen as relevant to the broader domain of prototyping Human-Computer Interactions. As an important goal of our own research was to support exploration of multilingual scenarios, we focus on systems that may be capable of integrating machine translation, although from the literature we can clearly see that WOZ is predominantly used in monolingual settings.

% =============================================================================
\section{Wizard of Oz and its applications}
% =============================================================================
Human simulation as a prototyping method was first applied more than 40 years ago when~\cite{Erd71} tested their concept of a self-service airline ticket kiosk and then later when~\cite{Gou83} explored the possibilities of the `Listening Typewriter'. The name `Wizard of Oz' or `OZ Paradigm', respectively, was given to the method by~\cite{Kel83} who used it to simulate a calendar application that could be operated via natural language input. Thereafter several researchers employed this new technique for prototyping natural language based interaction \citep{Goo84,Gou87,Car88,Hil88,Jon88}, which sometimes was also referred to as PNAMBIC (Pay No Attention to the Man BehInd the Curtain) \citep{Fra91}, before first~\cite{Hau89} and then later \cite{DeM93} extended its application area from testing purely text- and speech-based interaction to evaluating gestures and face recognition.

This expansion in scope continued with \cite{Sal93} who looked at multi-modal interaction, leading to the introduction of multiple wizards. In more recent years WOZ experiments have been used for a variety of purposes, including prototyping multi-modal information retrieval \citep{Raj06}, testing speech-based flight booking systems \citep{Kar08} and simulating a virtual doorman \citep{Mak01}. Exploring relatively open interaction spaces, \cite{Bra09} used WOZ to evaluate users' experiences when interacting with a web-based social companion, \cite{Gol99} employed it to investigate navigation in voice-controlled dialogues, and \cite{Dav98} tested the advantages of active help when using an unfamiliar software application.  Further examples of WOZ experimentation and how they are used can be found in~\cite{Dah93}. 

While these examples illustrate a variety of use cases for the method, the vast majority of them fit the categories described in the preceding section, namely: exploring interaction strategies \citep{Okamoto01} and designing dialogues \citep{How05}, collecting text and speech corpora \citep{Ben03}, and evaluating components \citep{Skantze10}. In terms of its use in interaction design, as with low-fidelity prototyping methods for software based on GUIs, WOZ can play a role in shaping an application structure and improving the `naturalness' of an interaction. The method supports designers in producing appropriate dialogue models and allows them to improve their understanding of a domain.

Although user behaviour is usually the focus of WOZ studies, analyses of wizard behaviour have also been conducted, for example in a study by \cite{bib:RieserLemon10l}, who used a WOZ setting to gather data on whether to present visual information or to use only speech in clarification requests. A final area in which WOZ was found to be helpful is the exploration of emotions \citep{Sch08} and social aspects of human-machine interactions \citep{Der05}.

% =============================================================================
\section{Wizards as Users and their Interest in the Method}\label{sec:users}
% =============================================================================
%From a system design perspective, we can generally find two distinct groups of users for the WOZ method and its associated tools. 
From a system design perspective, WOZ is unusual in that two distinct groups could be regarded as its potential users. On the one hand, we may identify the prospective end-user of the prototyped system as the main party involved in the method. However, as the WOZ technique mainly enables a designer to explore a variety of application areas, designers and researchers themselves can be seen as the more central user group for a tool. Their task (to design and run WOZ experiments) is complex and their general requirements poorly understood. In this paper we therefore focus on the person(s) who design and run WOZ experiments as the immediate user group rather than the potential end-users of a prototyped system.

From a design perspective, students studying Human-Computer Interaction and Interaction Design will generally be introduced to WOZ, yet only a small proportion of these will actually experience the method when compared to exercises based on the use of paper prototypes. One reason for this lack of practical usage might be that in order to be applicable in an HCI teaching context, any approach would have to have a low logistical and technical overhead to enable students to quickly design and carry out evaluations. 

Experienced interaction designers are another obvious user group. Our own interviews with developers of systems based on Interactive Voice Response (IVR) suggest that WOZ is used sporadically within product development. However, the opinion was expressed that the limited time-scale typically available for interaction design activities, especially within smaller projects, often impedes the application of the method. Hence, it seems that more exploratory uses, such as those represented by the HCI research literature, may be a more sensible starting point, when it comes to understanding users coming from the area of voice-interface design and development.
%Hence, it seems that a broad range of more exploratory uses, such as those represented by the HCI research literature, would be a more sensible starting point, as opposed to some specific sub-category of users who make heavy use of language technology.

Another distinct category of users are people working in computational linguistics as they usually have a strong interest in gathering language corpora. Such corpora are vital resources both for scholarly work and for the development of Language Technology Components. For example, the WOZ method can be used to collect an initial language corpus upon which components are trained and improved \citep{Lam98}. Collecting context-specific and language-specific corpora helps to expand the reach of existing technologies. The desired output from an experiment in this setting is typically the input supplied by the non-wizard user, whether it is typed text, speech, or multi-modal input (for example speech and gestures).

Those involved in the development of these technology components may also be interested in WOZ as it allows them to evaluate the performance of their products in a real-world setting, for example within a specific application context. Using WOZ these technologies can be tested in more realistic, task-focussed evaluations (many existing language technology component evaluations are otherwise based on context-free standardised benchmarks), without the need to construct a fully working system around them (which is usually not the focus of their work). The usual benchmark of such an experiment might be the word error rate for the recognition of application-specific utterances, rather than the design of the dialogue itself or other aspects of task performance.

Finally, while we do not focus on it in this paper, WOZ has also been used in psycho-linguistic research into how human-human dialogues differ from human-computer dialogues. In particular the area of syntactic and lexical alignment has been the focus of recent work~\citep{Bra03, Bra11,Cow12}.

Looking at these different user groups and their interest in employing the method one can find again the three main application areas for WOZ experimentation: Firstly, interaction design, where the flexibility to explore a range of different types of scenario is key, but which might make use of language technology components as part of delivering an authentic experience. Secondly, component evaluation, testing the quality of existing technology, which requires that (at least partially) working components be integrated, and finally corpus gathering, which may or may not require the integration of working components.

% =============================================================================
\section{Learning from experiences with Wizard of Oz}
% =============================================================================
In order to further increase our understanding of the different user groups and their distinct usage scenarios for WOZ we conducted an interview study with researchers from industry and academia. In total five professional voice interface designers and 25 academics who had recently published relevant work in the area (i.e. mostly within the last five years) were approached via email and asked for a phone interview. Positive responses from three of the designers and 17 researchers (seven of them working in NLP, five in HCI, and five in the area of multi-modal interaction) led to a total of 20 interviews, each of which lasted between 17 and 30 minutes. All of the interviewees were actively involved in at least one WOZ study, and 13 of them indicated that they had used WOZ in a variety of experiments. Interviews were semi-structured and participants were asked about their motivation for using the method, the challenges they had to overcome when doing so, and the tools they had employed. The recordings were fully transcribed and analysed through an open coding process before grouping them inclusively in the sections highlighted below (results are summarised in Table~\ref{table:interviews}). The coding and thematic analysis was carried out by one of the authors and subsequently cross-checked by another member of the team.

\begin{table*}
\processtable{Summarised results of 20 phone interviews conducted with researchers from industry and academia who have experience with WOZ experimentation.\label{table:interviews}}
{\begin{tabular*}{\textwidth}
{l | l}
\toprule
\multirow{4}{*}{Reason for using WOZ} 
& Explore new design ideas\\
& Collect an initial dataset\\
& Compare specific design solutions\\
& Evaluate technology \\
\midrule
\multirow{7}{*}{Challenges to overcome} 
& Delays caused by aspects of the wizard task\\
& Make participants believe that they are interacting with a system\\
& Simulate consistent system behaviour\\
& Simulate erroneous or suboptimal system performance\\
& Recruit participants\\
& High experiment costs\\
& Ethical issue of deceiving participants\\
\midrule
\multirow{1}{*}{Tools employed} 
& Self-developed programs supporting a very specific experiment setting\\
\botrule
\end{tabular*}}{}
\end{table*}

\subsection{Reasons for using WOZ}
Exploring new design ideas before they are implemented was cited as a reason for using WOZ by the majority (13) of interviewees. This rationale is expressed in the following statement:
%A typical statement to that effect was S10: 
\begin{quote}
{\it So we had this idea of building this multi-lingual translation system but we were not very sure, so we wanted to do a WOZ simulation in which we placed a tri-lingual person in the middle.} (Participant S10)
\end{quote}
 The method was found to be especially useful as a means of obtaining early feedback on a proposed design direction (S03: ``\textit{So whenever you already have you know a draft design and you want to show it to customers or you just want to show it to an initial set of users in order to design it in the right direction.}'') or employ it as a low-fidelity proof of concept study (S10: ``\textit{Yeah this is what, this was just a proof of concept}'').

In addition interviewees, especially from the NLP domain, stressed the value of WOZ for collecting data (stated by 8 interviewees):
\begin{quote}
{\it You generate data without having a dialogue system and you create from this small dataset, you create simulated environments, and in that simulated environments you can train dialogue strategies.}(Participant S12)
\end{quote}
This was typically done in order to explore dialogue strategies (S09: ``\textit{Yes the research was focused on the dialog between the participants, the communication, what they would say.}''). Two researchers specifically highlighted its qualities for comparing specific design solutions (S14: ``\textit{The biggest point is to save time in developing the actual technology, to allow you to test out alternatives without over-committing to one of them early on.}''), in which case the possibility for quickly putting together different design proposals is a key property of the method:
\begin{quote}
{\it Like if you were thinking about a A or B design you can quickly put both together and then ride through you know half a dozen people and see which of the two designs seems to work better.}(Participant S03).
\end{quote}
 Finally, two others mentioned that they had used it to evaluate some of their technology components (S15: ``\textit{We performed WoZ experiments three times to evaluate our dialogue system.}'').

\subsection{Challenges to overcome}
In terms of problems researchers were facing it seems that delays coming from the wizard constitute the biggest challenge, specifically mentioned by 9 of the 20 interviewees (S07: ``\textit{The delay seemed to be the biggest problem.}''). These delays were attributed to a number of different aspects of the wizard task (S09: ``\textit{All they needed to do is type in a message and type enter.}''), information overload (S04: ``\textit{Moreover the problem was that theoretically I should only look at the non-verbal behaviour and the acoustic information.}'') or simply rooted in a lack of wizard training (S19: ``\textit{No, no specific training at all, we made some pilots.}''). Another particular challenge was found in `hiding the wizard' (mentioned by 7 interviewees) so that people would believe that they are in fact interacting with a piece of technology rather than a human being (S11: ``\textit{You have to make sure that users really don't feel that there is someone staying in the other room.}''). This requirement for realism of the simulated functionality is not restricted to giving the user the impression that they are interacting with a real system but it also involves reflecting the complexity of the underlying technology in a way that conforms to the designer's expectations (S20: ``\textit{The more complicated the technology the bigger the challenge of making the simulation reflect what might really happen.}''). 

Also in connection with this issue, some interviewees highlighted the challenge of maintaining consistent wizard behaviour (mentioned by 7 interviewees) and remarked on how inconsistencies can influence evaluation results (S16: ``\textit{I mean it has to be consistent. It has to give the same answer all the time.}''). This seems especially true for the quality and validity of the responses given to users, as variable wizard actions can lead to confusion for a test participant:
\begin{quote}
{\it  If you are not consistent then the user will be very confused by what they are seeing and they might give you feedback on something, on, well they will give you feedback and you, it will be hard for you to know whether or not they are responding or I should say which version of the interface they are responding to.} (Participant S18)
\end{quote}
Finally, the simulation of errors or suboptimal system performance was found to be important (mentioned by 2 interviewees), both for testing error-recovery routines as well as for conveying realistic system behaviour:
\begin{quote}
{\it So you start becoming better at mimicking a real system so from time to time you would throw in an error or a misrecognition or something that would basically make the participant to try to recover.} (Participant S03).
\end{quote}
 It can be seen as a way to reduce a wizard's workload while at the same time increasing the validity of the simulated system:
\begin{quote}
{\it  So I think that idea of we are going to just have the wizard do a very well defined task, so that they can focus on just doing that right and let the system simulate the errors and simulate the delays and simulate all that stuff.} (Participant S20)
\end{quote}
However, support for this sort of functionality is not generally available. Other challenges that were mentioned include the recruitment of participants (S09: ``\textit{And it took quite a long time to find participants, because it was very important that the participants, that both participants come.}''), the high cost of experimentation (S15: ``\textit{No special challenges, but, WoZ experiments cost too much. It is a big problem for us.}'') as well as the ethical issue of deceiving participants (S19: ``\textit{Yeah. You have to, you have to handle that with care. So there is some ethical, ethical considerations also to be made.}''). 

While recruitment is a common experimental problem, it can be argued that web-based approaches (such as those employed in remote usability testing) have the advantage of widening the potential pool of participants. With regard to cost, reducing the amount of technical effort required for constructing experiments, and reducing the logistical overhead of running and analysing experiments would be separate dimensions. Ethics is an important issue for WOZ as a methodology, as many experiments will involve deceiving participants, and as such it is vital to ensure that participants are debriefed appropriately.
\subsection{Tools employed}
When asked about the tools they used to conduct their WOZ experiments, all of the researchers stated that they had employed self-developed programs and even though they mostly found that the implementation time for those solutions was feasible, the effort often appeared disproportionately high given that WOZ is usually regarded as being a low-fidelity prototyping method (S12: ``\textit{I think to develop a stable version of that, took us one person month at least.}''). Likewise, several interviewees expressed an interest in a more general WOZ prototyping tool, that could be adapted to their research interest in a flexible manner:
\begin{quote}
{\it Yeah, yeah it is very cool this idea, if it is researchers like myself ... that we can just manipulate, make it our own ... fit it to our own research ... and not having to develop a system on our own every time} (Participant S09). 
\end{quote}
Similar demands have emerged from within our own research environment. As part of an extensive research program on development and application of language technologies, we were increasingly facing the problem of how to test technology components with real users, without the overhead of creating an application environment for each case. Examples to that effect include the exploration of how MT technologies may be used in the work place~\citep{bib:DohertyKaramanisLuzCSCW2012,bib:KaramanisLuzDoherty11tpw}, how combining MT with other components such as TTS may influence the perceived user experience of products~\citep{bib:SchneiderLuz11salsm}, or how ASR might be used to help language learners better pronounce foreign words~\citep{Cab12}.   

\MNOTE{SS: I've slightly expanded this section to include some examples and references of the work conducted as part of CNGL}

These practical examples combined with the previously analysed literature and complemented by the interview study described above, supports the conclusion that there is a need for more generic tool support which to date has not been addressed or explored sufficiently, and that such tool support should pay particular attention to experimentation involving real or simulated language technology components.

% =============================================================================
\section{Requirements for Wizard of Oz Tool Support}
\label{sec:requ-wizard-oz}
% =============================================================================
Having started with the different groups of users possibly interested in employing the WOZ method, and their distinct application scenarios, we can now move on and look more closely at the requirements for tool support, what existing tools offer and where support should be improved. We can start by categorizing requirements for WOZ experiments into features and qualities (summarised in Table \ref{table:requirements}).

\begin{table*}
\processtable{General requirements for supporting WOZ experimentation.\label{table:requirements}}
{\begin{tabular*}{15cm}
{l | l}
\toprule
\multirow{3}{*}{Features} 
& Support both structured and flexible interactions\\
& Support integration of (unreliable) components with human intervention\\
& Support experimental data capture and export for analysis\\
\midrule
\multirow{2}{*}{Qualities} 
& Reduce overhead in experiment construction and software installation\\
& Reduce cognitive burden on Wizard during experimentation\\
\botrule
\end{tabular*}}{}
\end{table*}

Firstly, from a functional point of view a WOZ tool would need to provide support for running tightly controlled experiments as well as more exploratory studies. Even within tightly controlled experiments some flexibility for dealing with the unexpected may be useful. Features that would allow for highly structured interactions include the possibility for creating, selecting and grouping responses, and the availability of filters and similar aids that help a wizard retrieve information. 

Secondly, when we consider scenarios where different existing technologies are combined, there is a potential for this interconnection to increase failure. A representative example would be the analysis of speech-to-speech translation where the output of an ASR component is often used as input for MT and its output then fed to the text-to-speech synthesiser. Whereas humans might very well tolerate small mistakes coming from single components, technology is less forgiving. That is, while humans may use contextual information to handle small speech recognition errors, they can easily lead to problems when forwarded to a translation service. Supporting the function of a human in the loop who acts as an enhancement rather than a replacement for the technology would allow exploration of  these kinds of dependency problems in more detail.

Thirdly, tracking mechanisms and data exports would need to be available in order to analyse user behaviour. In addition, being able to gather data on wizard task performance and how it changes depending on the experiment setting, and over the course of an experiment, can be seen as a feature that could make this prototyping method more robust. Existing problems with this sort of data logging were explicitly mentioned in our interview study (S12: ``\textit{Actually the logging is another challenge which ... what happened to us is that we lost data}''), and therefore clearly highlight its importance and the need for improvement.  

On the other hand, from a more qualitative point of view we see that currently the requirements for installing multiple software components and configuring the network (to support the connection between the user and the wizard) quickly increases the amount of time and resources needed for running WOZ, and therefore  diminishes its value as a low-fidelity prototyping method. A further complication is that technology components are often platform-specific. Hence, reducing this cost of setting up, designing and running experiments would make the method more attractive and accessible to researchers and designers of all fields. 

Finally, another qualitative aspect that currently poses significant challenges for WOZ experiments, is the workload of the human wizard while running evaluations~\citep{Sal93}. Correct timing, consistency and general machine-like behaviour directly influences the validity and representativeness of an experiment, and therefore, if not controlled, can influence evaluation results. Support could come from visible instructions and reminders that might help the wizard achieve consistency, or from highly customizable wizard interfaces. Furthermore, if we consider changing the wizard's role from replacing to enhancing technology, as outlined above, additional support for that role might also be required. 

% =============================================================================
\section{Existing Wizard of Oz Tool Support}
% =============================================================================
Even though there seems to be a clear demand for integrating WOZ support into language technology frameworks, only a limited range of applications offer adequate functionalities to do so. From the literature, the software tools and frameworks that have been used for prototyping language-based interaction scenarios differ greatly between the different application scenarios. Furthermore, many of those referred to require a considerable amount of set-up time and often depend on obsolete technology. Many also do not appear to be publicly available.

Generally applications and frameworks that may support WOZ exploration can be separated into two categories. In the first category we find Dialogue Management (DM) tools which focus on the evaluation of language technology components and whose primary application lies in the area of Natural Language Processing (NLP) and machine learning. Tools from the second category, herein referred to as pure WOZ tools, instead rely completely on human simulation, which makes them more suitable for exploratory analyses.

% =============================================================================
\subsection{Dialogue Management Tools}
% =============================================================================
Two of the better known examples for DM tools are the CSLU toolkit \citep{Sut98} and the \textsc{Olympus} dialogue framework \citep{Boh07}. Others include the \textsc{Jaspis} dialogue management system \citep{Tur00} and the EPFL dialogue platform \citep{Cen05}. DM tools explore the language-based interaction between a human and a system and aim at improving this dialogue. They usually provide an application development interface which is used by a programmer to specify the dialogue flow and its integration of different language technology components like ASR and TTS. Once designed the dialogue is tested using human participants. In doing so the main focus lies on testing and improving the quality of the technology components used. Typically, these tools depend on the language technology components that are integrated which means that test results will depend heavily on the quality of the existing technology. %Only crude support is available for human intervention in the form of WOZ.

The CSLU toolkit for example, offers speech-recognition, natural language understanding, speech synthesis as well as a talking head. Modules are integrated into a stand-alone graphical authoring environment, which allows  dialogue flows to be specified. Dialogue elements are dragged onto a canvas where they can be arranged and linked using a flow chart-like notation that also supports decisions, random generators and loop backs. Input and output can be defined separately for each element so that it is possible to integrate and combine text, spoken and touch-tone based interaction. Even though an integration of WOZ support was planned, to our knowledge the functionality never made it into any of the final product releases. In general, however, the CSLU toolkit can be seen as a straight-forward prototyping tool that requires little experience, which makes it suitable for both designers as well as NLP researchers. Providing a simple graphical interface increases accessiblity for people without a technical background. 

In contrast, the \textsc{Olympus} dialogue framework constitutes a powerful client-server environment for implementing and running spoken dialogue systems. The goal of the framework is to provide a highly scalable platform for language technology research, yet its support for quick prototyping is low. Composed of several different components (i.e. an audio server, the Apollo interaction manager, the Phoenix grammar parser, the RavenClaw dialogue manager \citep{Boh03}, the Rosetta language generator, and the Kalliope speech synthesiser), none of which provides a graphical interface, it requires a high level of technical know-how to set-up and use. Also, while WOZ experimentation is certainly possible \citep{Boh05} it requires the relevant component to be built on a one-off basis and integrated with the rest of the framework. A ready-made WOZ client is not available. Nevertheless, this loose coupling of different technology components allows for a high degree of flexibility, which makes the framework suitable for evaluating new technologies. 

Similarly, the the \textsc{Jaspis} dialogue manager provides high adaptability. While predominantly aimed at building working systems, it puts a strong emphasis on information representation. The XML-based output allows for integrating natural language into applications that run on different devices, representing a range of form factors \citep[e.g.][]{Tur05}. While here also WOZ experiments have been conducted \citep[e.g.][]{Mak01}, the integration was achieved through building a one-off interface rather than integrating generic WOZ support. 

Finally, the EPFL dialogue platform shows some more general support for WOZ integration with language technology. Based on the Rapid Dialogue Prototyping Methodology (RDPM) it automatically creates a graphical wizard interface based on a pre-defined application model \citep{Raj06}.
This automatic generation of interfaces makes the platform interesting for researchers that have little technical knowledge. In addition it supports multi-modal as well as vocal designs, extending its application domain  beyond purely language-based interaction paradigms. 

In summary, existing DM tools highlight several important requirements for supporting the prototyping of language-based applications. Graphical, stand-alone tools like the CSLU toolkit provide a low entry barrier for non-experts.  On the other hand we see that a high degree of component flexibility, as demonstrated by the \textsc{Olympus} dialogue framework, opens up a wider range of possibilities, especially when it comes to evaluating different technological solutions. Also, support for new form factors, as can be found with the XML-based architecture of the \textsc{Jaspis} dialogue manager, seems crucial, particularly when we think about mobile phones, tablet computers and their potential successors. Finally, the dynamic generation of interfaces, whether for wizard or for tested client interfaces, is a feature that helps to significantly reduce the prototyping time. 

While the described examples might not cover the totality of DM tools that have been used in the past, they all show that WOZ prototyping has its place in dialogue design - although generally additional development work to produce the relevant interfaces is required. %Combining some of the features presented with a better WOZ integration might therefore create a new class of product; one that is both flexible and easy to use without requiring a significant amount of integration effort.

% =============================================================================
\subsection{Pure WOZ Tools}
% =============================================================================
Unlike DM tools, pure WOZ tools try to more fully support low-fidelity prototyping. While these applications offer more flexibility than DM tools, they usually do not integrate actual working language technology components. Instead a human mimics the functions of the system, which allows for a less restrictive dialogue design. In addition it facilitates the testing of user experiences that are not yet supported by existing technologies. Pure WOZ tools are, however, scarce and tend to be only suitable for the one experiment for which they were constructed. Hence, they are often categorized as throwaway applications i.e. they are built for one scenario and only rarely re-used in other settings. Two publicly available tools that allow for more generic experimentation include SUEDE \citep{Kle00} and Richard Breuer's WOZ tool\footnote{http://www.softdoc.de/woz/index.html [Accessed:  4$^{th}$ Sept. 2013]}. An alternative, yet not publicly available, solution may be found in the \textsc{NEIMO} platform~\citep{Cou96}.

SUEDE allows a designer to rapidly create prompts and supports the graphical design of a dialogue flow. It lets the researcher record those prompts, arrange them, and play them back to a test participant. During experimentation a participant's responses can be captured and the collected data be made accessible in form of a browsable HTML document, which can be used as a reference for future design improvements. The advantage of this type of evaluation is that a designer can focus entirely on the interaction, while keeping the quality of the speech output (i.e. the pre-recorded prompts) consistent; something that might vary if actual technology components are used. The exclusion of third party components further reduces the complexity of a test set-up which ultimately leads to less time spent on the configuration of experiments. Two main features of general interest in SUEDE are first the provision of a simple way of recording, organising and playing back potential system prompts, and second structured access to participants' responses, which is crucial for understanding and consequently improving the overall interaction.

Following a similar goal, namely supporting simple, generic WOZ experimentation, Richard Breuer's WOZ tool puts a stronger focus on more complex dialogue designs. It does not offer a dedicated function to record prompts, however, one can link dialogue elements to stored audio files or make use of an integrated text-to-speech function. Furthermore, the tool allows dialogue flows to be exported as VoiceXML\footnote{http://www.w3.org/TR/voicexml21/ [Accessed: 23$^{rd}$ Dec. 2013]} or RTF, so that they can be re-used in third party products. This support of XML standards is an important feature that helps to integrate WOZ experimentation as a fundamental part within the development cycle of new dialogue systems.  

Finally, a third application, the NEIMO platform, was mainly used in the 1990's to study the potential of multi-modal user interfaces \citep{Bal93}. The demand for evaluating this sort of interaction has significantly increased since then, and so it is clearly useful to consider support for multi-modal interaction when developing prototyping tools. One important finding with respect to WOZ prototyping gained from these studies was that additional modalities also significantly increase a wizard's workload. Hence, NEIMO was one of the first tools to support multiple wizards. Roles could either be split up between the different modalities or were dedicated to the input/output interpretation/generation on the one hand and task level processing on the other. 

The feature set of the pure WOZ tools discussed above ranges from simple graphical interfaces for designing a potentially multi-modal interaction, to powerful logging and export functionalities that make use of industry standards such as VoiceXML and therefore smooth the path to a possible integration with external systems. Combining these features seems to be the logical next step towards better WOZ support.  

However, if we look at more recently created WOZ tools (also outside the area of language technology applications), we find less generic support but rather a focus on very specific application scenarios. Hence, a high level of the development effort is required to adapt the tools to different domains. Examples include \textsc{MDWOZ} \citep{Mun00} for dialogue systems, \textsc{QuickWoZ} \citep{Sme10} to study embodied conversational agents, \textsc{WOZ Pro} \citep{Hun07} and \textsc{SketchWizard} \citep{Dav07} for simulating pen-based interaction, and \textsc{Polonius} \citep{Lu11} and DOMER \citep{Vil11} to control a robot. WOZ functionality can also be found in \textsc{Topiary} \citep{bib:LiHongLanday04t} and \textsc{BrickRoad} \citep{bib:LiuLi07b}, tools for prototyping location-enhanced applications. Finally,~\cite{bib:LiuRieserLemon09} describe a WOZ interface to support the study of information presentation strategies for spoken dialogue systems and,~\cite{bib:OttoFriesenRoesner11mozh} developed a tool based on the \textsc{{SEMAINE}} framework\footnote{http://semaine.opendfki.de/wiki/SEMAINE-2.0 [Accessed: 23$^{rd}$ Dec. 2013]}, a multi-modal dialogue system that aims at sustaining conversations with human users. 

An exception to these rather specialised tools can be found in \textsc{Ozlab}~\citep{Pet02}, a multi-functional prototyping tool for multiple forms of WOZ experimentation. Here the authors explicitly highlight the re-usability of their tool. However, we were unfortunately not able to evaluate the application in more detail, as it is not publicly available. Finally, focusing on information retrieval tasks,~\cite{bib:SchererStrauss08fwozen} present a flexible WOZ environment for spoken dialogue systems that makes use of other parallel output modalities (e.g. a talking head). But again, an in-depth analysis was not possible due to the lack of availability.

In summary, the majority of existing pure WOZ tools either suffer from dependencies on obsolete software components and accompanying compatibility problems  or pose considerable challenges when there is a need to adapt them to new application scenarios. Also, they are rarely publicly available, which restricts their re-use by other researchers. In addition, a potentially problematic issue with most of the tools is their shift from relying completely on technology to relying completely on the actions of a human. These are both extremes on what we can see as a continuum, where the intervening points represent a mixed-fidelity approach in which imperfect components can be incorporated along with human intervention. 

% =============================================================================
\subsection{Challenges of Generic Tool Support}
% =============================================================================
While the above analysis discussed the two different forms of WOZ support that may be available for language technology applications (i.e. support through existing wizard functionalities in dialogue management tools and also support through dedicated WOZ applications) one may also conclude that their efficient integration in the product development cycle is difficult. Support for this argument can be found in the fact that applications that support the simulation of an interaction are commonly seen as throwaway tools which are only used for a certain number of experimentation rounds. While efforts have been undertaken to increase the re-usability of applications with some researchers going as far as aiming for providing a generic evaluation platform, the distinct requirements of different test scenarios often require significant changes to be implemented. Within the same research team these changes are usually implemented, which leads to tools that try to be generic, but to some extent include features that are very specific to a certain application area.

When it comes to using the tools of third party providers, adoption can be inhibited by a lack of technical documentation or simply because of the amount of time that is needed to become accustomed with the relevant code base, which may be comparable to what it would cost to build a separate tool. We mainly see this when looking at the variation of pure WOZ applications presented in the literature. For example, \textsc{Polonius}~\citep{Lu11} and \textsc{DOMER}~\citep{Vil11} are both wizard tools that are used to control a robot. While the underlying research interest for which they were built might differ, they share the same core functionality, namely sending essentially pre-defined commands to a remote system.

One can also see similarities with \textsc{QuickWoZ}~\citep{Sme10}; although the focus of \textsc{QuickWoZ} lies on avatar-based interaction, the underlying concept remains the same. From an application point of view, the difference between sending commands to a robot and controlling the feedback given through an animated character on a screen is relatively small. Similarities can also be found between \textsc{MDWOZ}~\citep{Mun00} and \textsc{DiaWoZ}~\citep{bib:FiedlerGabsdil02swoz} as well as between \textsc{WOZ Pro}~\citep{Hun07} and \textsc{SketchWizard}~\citep{Dav07}. An obstacle for re-using a tool may, however, be found in the programming framework that is used to build it, although in this case a possible solution can be the provision of appropriate software interfaces.

If we look at dialogue management tools, generic support seems more complicated. In this case, as WOZ is often treated as an integrated function rather than being a separate external tool environment, the potential re-usability is limited to the re-usability of the entire dialogue framework. Here WOZ changes from being an evaluation method informing the design to being an integral part of the final product. Nevertheless, by focusing on a modular composition as demonstrated by \textsc{Olympus}~\citep{Boh07}, the WOZ function could be liberated and consequently re-used with other comparable frameworks. An advantage of this separation process is not only that a dedicated WOZ module could be integrated with other dialogue environments, but it would also allow WOZ to be treated as as an independent component whose further development could be influenced by multiple parties inside as well as outside a given research team. External feedback would also make it easier to improve and fine-tune existing functionalities as well as allow for gradually integrating novel use cases inspired by new application scenarios.  

While interoperability between WOZ tools seems desirable both to save resources spent on building proprietary solutions and to expand rather than re-build already existing functionality, the applications that have been published in the literature show that the exchange between research teams is rather limited. Despite the fact that numerous examples advertise their high flexibility and easy reconfigurability so that in theory re-use of applications would be possible, researchers solve their specific problems by creating new tools rather than improving existing ones. One reason for this is surely to be found in the varying research interests and their very distinct requirements when it comes to tool support. Another aspect is that for applications outside the NLP domain WOZ often plays a minor role for which a quick-and-dirty solution is usually sufficient and building re-usable components might seem unreasonable.

Furthermore many tools described in the literature are not freely available, or suffer from dependencies on obsolete software components. Insufficient documentation is a further barrier to adoption and reuse. \textsc{SUEDE} and Richard Breuer's WOZ tool are the only pure WOZ tools which we were able to download. With dialogue management tools we were restricted to the~\textsc{CSLU} toolkit and the \textsc{Olympus} dialogue framework. Finally, little effort has been put into understanding and improving WOZ prototyping to the point where it can be treated as a separate area of competence. After all, for most research teams WOZ is not an end itself, but rather a method to evaluate and improve the design of an envisioned future product. To do so, they come up with solutions that mainly serve their very specific requirements. Tools whose goal is to provide improved generic support for the WOZ method are, however, missing. 
% =============================================================================
\section{The Task of the Wizard and its Design Space}
\label{sec:designspace}
% =============================================================================
While DM tools and pure WOZ tools both incorporate useful features, neither type of tool provides a full range of support for the use of WOZ as a low-fidelity prototyping method. In order to build more appropriate and potentially generic instruments we need to better understand the task of the wizard as well as the design space for WOZ.

\begin{figure*}
\centering
\includegraphics[width=0.7\textwidth]{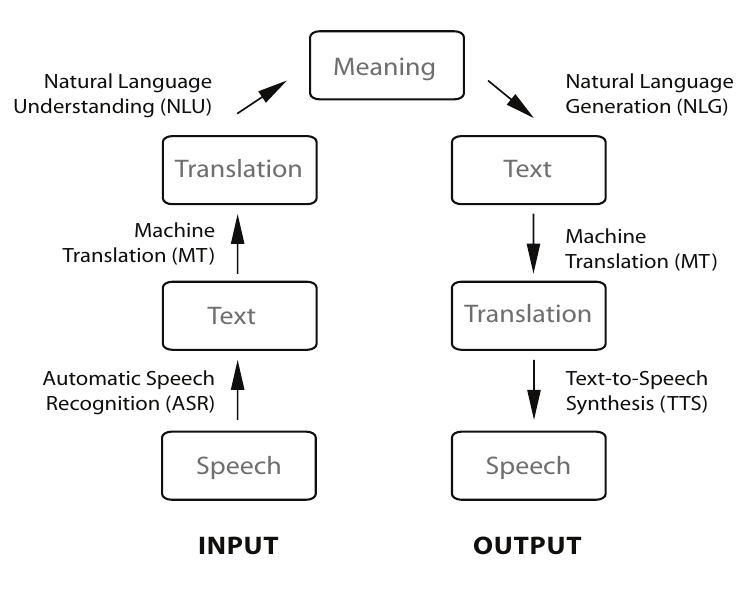} 
\caption{The Interaction Pipeline of Language Technology.}
\label{fig:interactionpipeline}
\end{figure*}

We start our exploration of this domain with a consideration of available language technology components and how they might be integrated. Researchers have reported on WOZ evaluations in the area of Human-Machine dialogue as well as computer supported Human-Human dialogue. The latter is especially relevant to machine translation where technology aims to build a bridge between people who do not share a common language \citep{Bed10}. From a more component-based view, WOZ has been used to simulate ASR, MT, Natural Language Understanding, and Natural Language Generation as well as TTS. Only rarely has it been used to enhance existing technology (i.e. to correct or over-write what a semi-working technology component might output). Examples include the adaption of a storyline \citep{Dow10}, the mimicking of social behaviour \citep{Der05}, or the annotation of language \citep{Jan09}. 

From a Natural Language Processing (NLP) point of view, one can envisage a generic pipeline architecture that starts on the input side with Automatic Speech Recognition (ASR) and ends on the output side with Text-to-Speech Synthesis (TTS). In between we might further find Machine Translation (MT) on either side of the dialogue management component (see Fig. \ref{fig:interactionpipeline}). In non-speech scenarios ASR and speech output can be replaced by other, text-based, input and output modalities. These 4 dimensions of input modality, input MT, output MT, and output modality lead to a total of 16 different possible task settings a wizard might have to deal with when running WOZ experiments (see Table \ref{table:designspace}).

\begin{table*}
\processtable{Design Space for WOZ scenarios with language technolog components and associated application examples.\label{table:designspace}}
{\begin{tabular*}{15cm}	
{lcccccccm{4.2cm}}
\toprule
& \multicolumn{2}{|c|}{Input} & \multicolumn{3}{c|}{Processing} & \multicolumn{2}{c|}{Output} & \\ 
\cline{2-8}
Case & \multicolumn{1}{|c}{Text} &  \multicolumn{1}{c|}{ASR} & MT & DM & \multicolumn{1}{c|}{MT} & TTS & \multicolumn{1}{c|}{Text} & Example \\
\midrule
1 & x & - & - & x & - & - & x & Natural-Language Interfaces \\ 
2 & - & x & - & x & - & - & x & Speech Recognition \\ 
3 & - & x & x & x & - & - & x & Text-based Feedback \\ 
4 & x & - & x & x & - & - & x & Text-to-Text Translation \\ 
5 & x & - & - & x & x & - & x & Text-to-Text Translation \\ 
6 & x & - & - & x & - & x & - & Speech-output \\ 
7 & x & - & x & x & - & x & - & Multi-lingual Text-to-Speech \\ 
8 & x & - & - & x & x & x & - & Multi-lingual Text-to-Speech \\ 
9 & - & x & x & x & x & x & - & Less common \\ 
10 & x & - & x & x & x & - & x & Less common \\ 
11 & - & x & x & x & x & - & x & Less common \\ 
12 & x & - & x & x & x & x & - & Less common \\ 
13 & - & x & x & x & - & x & - & Speech-to-Speech Translation \\
14 & - & x & - & x & x & x & - & Speech-to-Speech Translation \\ 
15 & - & x & - & x & - & x & - & In-Car Voice Control \\ 
16 & - & x & - & x & x & - & x & Multi-lingual Inf. Retrieval \\
\botrule
\end{tabular*}}{}
\end{table*}

Usually the task of the wizard is to replace a single component or a combination of several. In most cases the language understanding and the generation of an appropriate response are simulated, supplemented by one or more additional components (e.g. ASR or MT). Therefore we may conclude that Dialogue Management (DM), Natural Language Understanding (NLU) and Natural Language Generation (NLG), are the main tasks a wizard needs to deal with. While often these three aspects are integrated into a single task in which the wizard extracts the meaning from an input and subsequently initiates the appropriate output, it should be noted that there are a number of test settings where NLU and NLG are broken down into separate components. Consequently the wizard may simulate only one of the two or simply act as a link between them. 

In some cases the resulting dialogue management task will be trivial (as in some real-time translation scenarios), in other cases complex `form-filling' steps may be required in order to progress. While in the following discussion we focus on cases where the dialog management is performed by the wizard, other scenarios might involve a (partially) working DM component. Looking at the combinations of technology components (working, partially working, or simulated) that might be combined in different scenarios, we obtain the design space shown in table \ref{table:designspace} (Note: [x] signifies that for the given use case the component (working, partially working/corrected, or simulated) is present ; [-] signifies that it is not).

In the most basic form the interaction on both sides is based on text (Case 1) and the wizard's potential task is limited to managing the dialogue (i.e. interpreting text input, dialog management, and generating text output). An application scenario for this pure form of WOZ can be found in prototyping a chatbot or a natural language user interface \citep{Kel84}. Replacing text input with speech input by adding a (potentially simulated) ASR component may change the task of the wizard from interpreting text to interpreting speech (if not correcting output from a speech recogniser) (Case 2). Even though the difference may seem small it can lead to an increase in cognitive load for the wizard as spoken text cannot be revisited later on, which might lead to performance problems especially when dialogue partners use long sentences. An example for this form of interaction can be found in prototyping dictation software \citep{Gou83}.

The complexity of the wizard's task increases even more in cases where an additional translation component is involved. The simulation here could happen from speech input, which needs to be first processed and then translated (Case 3). In this case the task of the wizard can be compared to somebody simultaneously translating from one language into another \citep{Stu06}.

If the scenario requires text-based input only, the task of the wizard may be reduced from translating from speech to translating from text (Note: from a wizard perspective the same reduction in task difficulty may apply if a working ASR component is available and the wizard can rely on its output), either on the input (Case 4) or on the output side (Case 5). 
%Alternatively the recognition could be handled by a working ASR component or excluded for text-based input, which reduces the task of the wizard from translating from speech to translating from text, either on the input (Case 4) or on the output side (Case 5). 
Application examples for this type of setting include a multi-lingual chat \citep{Che08} or a text-to-text translation system \citep{Bed10}. Looking at the output side we see a similar combination of possible components. For example, prototyping a text-to-speech function of a tour guide would require a wizard to operate TTS from input text (Case 6) \citep{Okamoto01}, and in cases where the system needs to be multi-lingual an additional translation task can be found either on the input (Case 7) or on the output side (Case 8).

Here we also find different task settings depending on whether the wizard simulates both ASR and MT, or one or both components are available. Even with working components the Wizard task may be quite different in these cases, as in Case 7 the wizard would see machine translated input and possibly correct it, whereas in Case 8 the wizard would see (and possibly correct) or interpret human input which afterwards is translated. 
%Also here (partially) working components may be available which leaves the wizard with different possible task settings. On the one hand both ASR and MT could be simulated on the other hand one or even both components could be available and the wizard would focus on the dialog management

The highest degree of complexity exists in situations where the application scenario comprises the complete interaction pipeline as highlighted in Case 9. Even though this is possible (as illustrated later in this paper), it seems less likely that a WOZ setting would make use of (or simulate) MT on both the input as well as the output side. For the same reason Cases 10, 11 and 12 have been less explored. However, if an interlingua or an intermediary natural language is employed in an interactive situation, such as speech-to-speech multilingual dialogue \citep{bib:LevinGatesEtAl98}, it is entirely possible that a wizard might be required to affect both sides of the MT process. This sort of intervention could take the form of correction from source language to interlingua and simulation from interlingua to target language, for instance, or simply involve simulation of the translation process through mediation between speakers without actual translation, as done by \citet{bib:LuperfoyMiller97uspcd}.

A more likely case, however, would be simulation or correction of a speech-to-speech translation system in which MT is used only on one side of the pipeline (Cases 13 and 14) \citep{Kra96,Kik03}. Taking away the multi-lingual aspect, the setting of an IVR system would reduce the wizard's task to understanding or correcting speech input and producing appropriate speech output, either directly (perhaps using some sort of distortion device) or indirectly by choosing from a set of pre-recorded utterances (Case 15). Application areas for this sort of WOZ prototype include in-car navigation \citep{Geu02} as well as transactions such as booking tickets \citep{Lam98,Kar08}. Finally, multi-lingual information retrieval using speech (Case 16) would require the wizard to first process a spoken request in one language and then provide appropriate information from multi-lingual sources \citep{Schn10}.

The configurations detailed above provide a broad coverage of WOZ scenarios involving language technology components. However, we note that using WOZ for simulating multi-modal interaction dramatically increases its application area and at the same time places even higher demands on the wizard \citep{Sal93b}. Here the aspect of processing information coming from different input channels and aligning the respective output has been the focus of recent research \citep{Mel06,Lee08,Ser10}.

In addition to the variety of tasks a wizard may potentially be confronted with along the language technology pipeline, it is also worth looking at the different ways language technology output might be simulated or corrected. In the literature we find a number of experimental differences. One commonly used set-up places the wizard consecutively after an existing component where he/she is used to selectively correct or overwrite component output~\citep{Kar08}. This solution is particularly used in cases where a technological solution is available but error prone and hence the research team may be interested in determining the level of accuracy needed to meet existing user demands. An alternative type of WOZ prototyping may use a wizard in parallel to a technology component in order to produce improved language technology components output. For ASR this type of setting could be implemented using an N-best list where the wizard is confronted with a number of possible recognition results to choose from, or has to enhance the scoring of correct or partly correct recognitions~\citep{McI99}. A similar set-up is possible for MT components. The data generated can then be used to tune the machine learning algorithms the language technology component is based on. 

A setting could also use a language technology component to `overwrite' a wizard response. The main reason for this type of application is the introduction of errors so as to produce a more realistic component performance~\citep{Gou83,Fos98}. Finally, one may also define different constraints for the wizard so that experimental designs can reach from using a wizard that is able to freely generate responses, either typed or spoken (using some sort of distortion device)~\citep{Ste89}, to cases where the wizard is restricted to a pre-defined set of possible utterances~\citep{Bra09, Schn10}.

Another possibility is to combine pre-defined utterances, concatenating them and filling in missing pieces manually~\citep{Cab12}. At this point it should also be noted that even though we generally use the acronym TTS to refer to the production of spoken system output, the term `speech generation' may be more suitable as this would also cover the use of recorded utterances as well as include specific markup language which might be employed to emphasise certain speech characteristics (i.e. pitch, prosody, etc.).   

% =============================================================================
\section{A Comprehensive Tool Architecture}
\label{sec:architecture}
% =============================================================================
Looking at the design space outlined above it appears that combining technologies (a) is already best practice in language technology systems design, (b) would allow WOZ to be used in multiple stages of an application's processing, and (c) is necessary to allow WOZ to be used throughout the entire application lifecycle. Consequently a tool that aims to more comprehensively support the application of the WOZ method would need to offer a way of combining existing technologies in a more flexible manner. Some of those technologies might be fully working; others might still be in an early development stage, and would rely on a wizard to raise their quality to an acceptable level. To allow this, a software architecture is required which supports a flexible use of technology. Ideally, this would provide a modular, `pluggable' framework that allows components to be integrated or replaced easily. From an architectural point of view, we need to define each of the components, the different stages of development they can be in, and their relationships to each other. 

\begin{table*}
\processtable{Some examples from the literature showing possible component/state combinations.\label{tab:componentrelationships}}
{\begin{tabular*}{15cm}	
{lccccccm{1.8cm}}
\toprule
& \multicolumn{2}{|c}{Input} & \multicolumn{3}{|c}{Processing} & \multicolumn{2}{|c}{Output} \\ 
\cline{2-8}
Example & \multicolumn{1}{|c}{Text} & {ASR} & \multicolumn{1}{|c}{MT} & {DM} & {MT} & \multicolumn{1}{|c}{TTS} & {Text} \\
\midrule
\citep{Kel84} & ON & OFF & OFF & ON & OFF & OFF & Correcting \\ 
\citep{Bed10} & ON & OFF & OFF & Simulating & Simulating & OFF & Simulating  \\ 
\citep{Gou83} & OFF & Simulating & OFF & Simulating & OFF & OFF & Simulating \\    
\citep{Geu02} & OFF & Simulating & OFF & Simulating & OFF & ON & OFF \\
\citep{Schn10} & OFF & Simulating & OFF & Simulating & ON & ON & OFF \\  
\citep{Kar08} & OFF & Correcting & OFF & ON & ON & ON & OFF \\
\botrule
\end{tabular*}}{}
\end{table*}

As regards the different task variations a wizard can take on within the interaction pipeline, one can define several different modes technology components can be in \citep{bib:SchloglDohertyEtAl10w}. A component can be relevant for a given setting, that is, it is needed and therefore needs to be represented in some form (e.g. ASR in a hands-busy-eyes-busy situation), or it is irrelevant, in which case it must be possible to turn it off (e.g. MT in a monolingual setting). In the case where a technology is needed, one can further distinguish between three different states. Under the best circumstances the technology is of production quality and therefore can be used in a black-box manner producing results either for the wizard or a test participant. On the other hand, if the performance of a component is not sufficient, a wizard's task could be to enhance the quality of the component. This type of scenario is particularly useful when the goal of an experiment is to investigate the improvement in quality that is needed for a technology to be acceptable, and therefore requires some sort of correction mode. Finally, in a setting where a component is needed but not available, it is usually the task of the wizard to completely simulate the missing functionality.

In summary, a comprehensive WOZ tool should enable a wizard to complement existing technology on a continuum by allowing her to simulate and correct technology, before finally using it as a black-box. Likewise~\cite{Dow05b} argue that a wizard might first take on the role of a `controller' who simulates technology. Then, in a second stage act as a `moderator' who approves technology output, before finally moving on to being a `supervisor' who only overrides output in cases where it is really needed.
 
By looking at these different modes and carrying them on to the language technology level it is possible to further deduce a set of rules that handle the relationship between consecutive technology components. The first rule defines a fully working component as a black-box for which it can be preceded as well as followed by components in any state. If a component is simulated by the wizard, however, it needs to be followed by a working component. In cases where two or more consecutive components need to be simulated, they merge into a single task for the wizard (e.g. simulated ASR followed by simulated MT).
%When a corrected component follows a simulated one, both components merge into a single simulation task for the wizard, as it seems defective to first simulate input for a component and then correct its output. DELETED: covered 2 sentences down.
When one or more simulations follow a correction, they all merge into an integrated simulation. Finally, a component can only be in correction mode when either its preceding component is fully working or when it receives its input directly from a test participant. Table \ref{tab:componentrelationships} illustrates some examples of possible component-state combinations and the related task of the wizard. It should be noted that the focus here lies on single wizard scenarios. Introducing several wizards would relax some of these constraints so that consecutive components could be simulated or corrected separately by different wizards. Integrating this set of rules into a software architecture should allow for a more flexible use of technology when running WOZ experiments.

% =============================================================================
\section{Wizard of Oz and the Web}
\label{sec:web}
% =============================================================================
An increasing number of traditional software applications are now offered in a web-based form, and the applications available are becoming more complex. The almost ubiquitous availability of high-speed internet has been an important factor, but also recent advances in web technologies have been critical in supporting this transition from locally installed software to cloud-based web applications. While some of the WOZ experiment environments presented in the literature \citep[e.g.][]{Tur00} were built to some extent using web technologies, the majority are based on conventional software tools. The lack of simple support for web-based speech input and output has been a major obstacle, leading to the use of locally installed software, with associated installation effort, software dependencies and compatibility problems.

Recent advances in web technologies, however, provide better support for dealing with speech. Modern web browsers are able to process audio and video data in real time and without the need for additional plug-ins. Upcoming web standards (i.e. the forthcoming HTML5 standard\footnote{http://www.w3.org/TR/html5/ [Accessed: 23$^{rd}$ Dec. 2013]}) go further by allowing access to computer hardware through the browser. These standards open up new possibilities for WOZ experimentation. It is now possible to integrate speech input and output into a web-based platform, which significantly reduces the setup requirements for an experiment environment. Furthermore, by using web services it is possible to build flexible tool architectures, such as the one presented above, in a way which allows components to be integrated and replaced easily and on-the-fly.

As well as removing problems associated with installation, there is also a benefit in terms of interoperability with other platforms i.e. it is easy to integrate WOZ experiments into existing web-based software environments. For example, if a new interaction modality for a web-based help system needs to be tested, a WOZ client can quickly be added to an already existing interface. From the point of view of the wizard, it is further possible to add additional information channels such as video of the user or location data, which allows the evaluation of not only speech but also multi-modal interaction. Finally, the possibility of running WOZ experiments on different platforms with different form factors (e.g. smartphones, tablets, media centres) represents another significant advantage that web-based solutions have over traditional software.  

% =============================================================================
\section{Investigating Tool Support}
\label{sec:tool}
% =============================================================================
If we look at the previously outlined design space for WOZ it seems that, (a) including the different configurations described in Table~\ref{table:designspace}, (b) allowing for a flexible integration of technology components as shown in Table \ref{tab:componentrelationships}, and (c) offering all interactions via web interfaces as discussed in Section \ref{sec:web}, constitute the main features of an advanced WOZ tool, where a human wizard can work together with (imperfect) technologies. In order to evaluate this assumption and furthermore explore additional aspects of wizard support we have undertaken a series of evaluations, whose results were used to inform the design of a generic WOZ prototyping platform for language technology applications. First a simple web-based prototype was used as a technology probe, allowing us to explore the context of use and its distinct requirements. 

The goal was to identify those parts of the wizard task that are challenging, independent of the specific experimental setting, and therefore might require additional generic support. Informed by the results of this first evaluation a more generic WOZ prototyping platform was built and employed to more thoroughly explore the challenges of constructing WOZ experiments. An evaluation comparing the experiments constructed by more advanced users with those built by students showed that users from both groups were able to produce viable WOZ experiments. The following sections will describe these two rounds of evaluation in more detail and highlight some of their results.
%Again an initial evaluation with a small number of wizards was used to identify general usability issues. Following this, a second study in which one wizard interacted with 17 users allowed us to identify a number of potentially problematic aspects with respect to wizard consistency over time. Finally, an evaluation of the experiment construction process was carried out in which 10 participants used a slightly improved version of the prototyping platform to design language-based WOZ experiments. Table~\ref{table:evaluation} lists the different prototypes and in which evaluations they were used.

%\begin{table*}[]
%\processtable{The tool prototypes that were used to explore different aspects of Wizard of Oz experimentation.%\label{table:evaluation}}
%{\begin{tabular*}{\textwidth}
%{l | l}
%\toprule
%\multirow{1}{*}{Initial Technology Probe} 
%& {Evaluation Round 1: Exploring Wizard Workload}\\
%\midrule
%\multirow{2}{*}{1$^{st}$ version of WOZ Prototyping Platform} 
%& Evaluation Round 2: Building a generic Platform for WOZ\\
%& Evaluation Round 3: Analysing Wizard Consistency\\
%\midrule
%\multirow{1}{*}{2$^{nd}$ version of WOZ Prototyping Platform} 
%& {Evaluation Round 4: Supporting Experiment Construction}\\
%\botrule
%\end{tabular*}}{}
%\end{table*}

% =============================================================================
\subsection{An Initial Prototype and Requirements Study}
% =============================================================================
First, inspired by the literature \citep{Stu06,Che08,Geu02}, different experimental scenarios for WOZ were explored. The goal was to find realistic settings in which a combination of various language technology components (ASR, MT, TTS) would be required. By doing so we were able to obtain a more applied view on the design space and could further refine the initial set of requirements. Paper sketches were then used to design a wizard interface that, although initially hard-coded for one specific WOZ experiment, would be applicable to a variety of scenarios. A prototype for such an interface was built using basic web technologies (i.e. HTML, PHP and CSS). 

The wizard interface was split into two areas; one showed the dialogue flow holding the defined dialogue utterances, the other displayed domain data i.e. data relevant to an implemented scenario. The dialogue flow was subdivided into different stages in order to decrease the amount of visible utterances at a time. It was possible to manually switch between dialogue stages by clicking on the respective links. The utterances to be used in a particular stage were highlighted on the screen. By using the appropriate utterances the wizard was automatically led through the dialogue. For dealing with misunderstandings recovery utterances could be chosen from a separate area of the interface. In order to help the wizard choose suitable utterances, a set of filters were automatically applied based on the utterances that were previously sent. In cases where a test participant would change her mind, a wizard could manually update those filters without going back in the dialogue.

Having designed this initial wizard interface, a specific WOZ experiment was implemented simulating the speech-based interaction between a German speaking customer and a system recommending products (in our case products were different types of Internet connection bundles). An initial set of dialogue utterances for this customer-machine interaction was defined and tested for accuracy and completeness using a chat tool. After that a set of realistic WOZ experiments were conducted.

% =============================================================================
\subsubsection*{Setting}
\label{sec:set}
% =============================================================================
In order to have a realistic setting for our wizard evaluation, we chose an in-house study related to machine translation that was conducted by a researcher in a computational linguistics laboratory. She was observed acting as a wizard over 11 sessions with different test customers. Test customers were international students who were told that they would be interacting with a prototype of a new adviser system that would understand spoken input. They were asked to complete two tasks with the system. First they needed to obtain information on an offer for pre-paid Internet, and after that they were asked to inquire about a land-line contract. They were told that the system could understand spoken input but would reply via text output on the screen. None of the test customers knew that they were interacting with a human until after the test was completed. Participation in the study was voluntary and compensated with a \euro 10 book voucher.

% =============================================================================
\subsubsection*{Wizard Observations}
% =============================================================================

Detailed logs of all user and Wizard actions were recorded automatically, together with screen capture and audio, and an interview carried out immediately after the experiment.
With respect to the wizard task, our evaluation identified two general aspects which were challenging for our wizard and therefore would require additional tool support. 

%%R code for this chart
%
% w <- read.csv('w01_text.csv', header=T)
% plot.new()
% plot(w$IQR.Value, type='b', axes= F, ylim=c(0, max(w$IQR.Value)*1.1), xlab='Experiment trials', ylab='IQR for time:word ratio in sec', pch=1)
% axis(1, row.names(w))
% axis(2)
% dev.copy2pdf(file='fig/wiztest.pdf', paper='special', family='Helvetica')
%

%\begin{figure*}
%\centering
%\includegraphics[width=0.7\textwidth]{fig/wiztest.pdf} 
%\caption{The wizard had difficulties estimating a test customer's reading speed. Whereas a value %of zero can be seen as consistent, the wizard showed variations (IQR values) between 1 and 2 %seconds in any of the 11 trial runs.}
%\label{fig:wizardperformance}
%\end{figure*}

Firstly, it was difficult for her to find the right utterances and deal with domain data, i.e. she had problems finding the information demanded by customers. Influenced by the general layout of our interface, she had difficulties switching her attention from one area of the screen to the other. The problem was observed mainly at the end of an experiment when she needed to select responses from the main dialogue flow as well as from the domain data area (i.e. the area that was holding the offers for different Internet connection bundles), which led to confusion and delays. Furthermore the dialogue flow itself caused some difficulties. Even though our wizard was involved in the experiment construction and therefore familiar with all the utterances, she sometimes had problems finding the appropriate response.

This specific problem of information retrieval under time pressure was foreseeable and so our prototype design tried to offer support by providing a filter function and automatically adapting those filters based on the dialogue progress. However, the latter seemed to confuse the wizard, as it was observed several times that she manually changed filter values even though there was no need for doing so. A post-test interview conducted after the first experiment run indicated that she felt lost and that by manually adapting the filters she tried to regain control. Hence, one could argue that a wizard interface, despite leading the interaction, needs to leave the wizard in control as any automatic support functionality (such as the filter mechanism) may lead to confusion and consequently further increase the already high cognitive demand for this task. This, however, also raises an interesting question with respect to methodology, namely who should act as the wizard. From the literature, it seems that the experiment designer will generally act as the wizard. While this could potentially be a source of bias, the experience of the evaluations would suggest that using other, more domain experienced, people as wizards is a strategy which will encounter difficulty in practice without considerable familiarisation and training.

The second interesting aspect of the wizard task we observed was an issue our wizard had with the timing. Since the simulated interaction was in a speech-in-text-out format (i.e. a test customer was able to talk to the system, the system response was, however, text-based), it was difficult for our wizard to estimate the time a customer would need to read a response utterance. This problem was observed several times, when she assumed a problem with a sent response and therefore sent an additional utterance in order to `check' on a customer's status. Usually, however, there was no problem with the sent utterance, but the customer was simply not finished reading. 

% The log files of the 11 sessions highlight that this issue remains over time i.e. it seems %independent of the wizard's steady familiarization with the given set-up, showing variations of %the wizard's estimations between one and two seconds for any of the trial runs. 
%
%Fig. \ref{fig:wizardperformance} shows the interquartile range (IQR) per trial of the time our %wizard estimated a test customer would need to read a word (Note: An IQR of zero would indicate %consistent estimations).

An analogous problem would be knowing when a pre-recorded or synthesised speech utterance has finished output. This shows that a lack of status information can influence the interaction and therefore reduce the reliability of the produced experiment result. The problem of these acknowledgement tokens or back-channels has also been discussed by Jurafsky and Martin \citep{Jur08}. We may therefore argue that additional status information, either visual or acoustic, can be seen as an important feature a WOZ tool should offer. 

\subsection{{WebWOZ} -- A Generic Platform for WOZ}
Informed by these evaluation results, a first version of the WebWOZ Wizard of Oz Prototyping Platform was built. The goal of this platform is not only to tackle the discovered problems with our previous prototype, but also to move away from a tool that supports a single WOZ setting to something that would more broadly support the application of WOZ with respect to language technologies. The software architecture described in Section \ref{sec:architecture} was implemented and several language technology components, i.e. one ASR component, one TTS component and two different MT components, were integrated using web services.

\begin{figure*}
\centering
\includegraphics[width=0.6\textwidth]{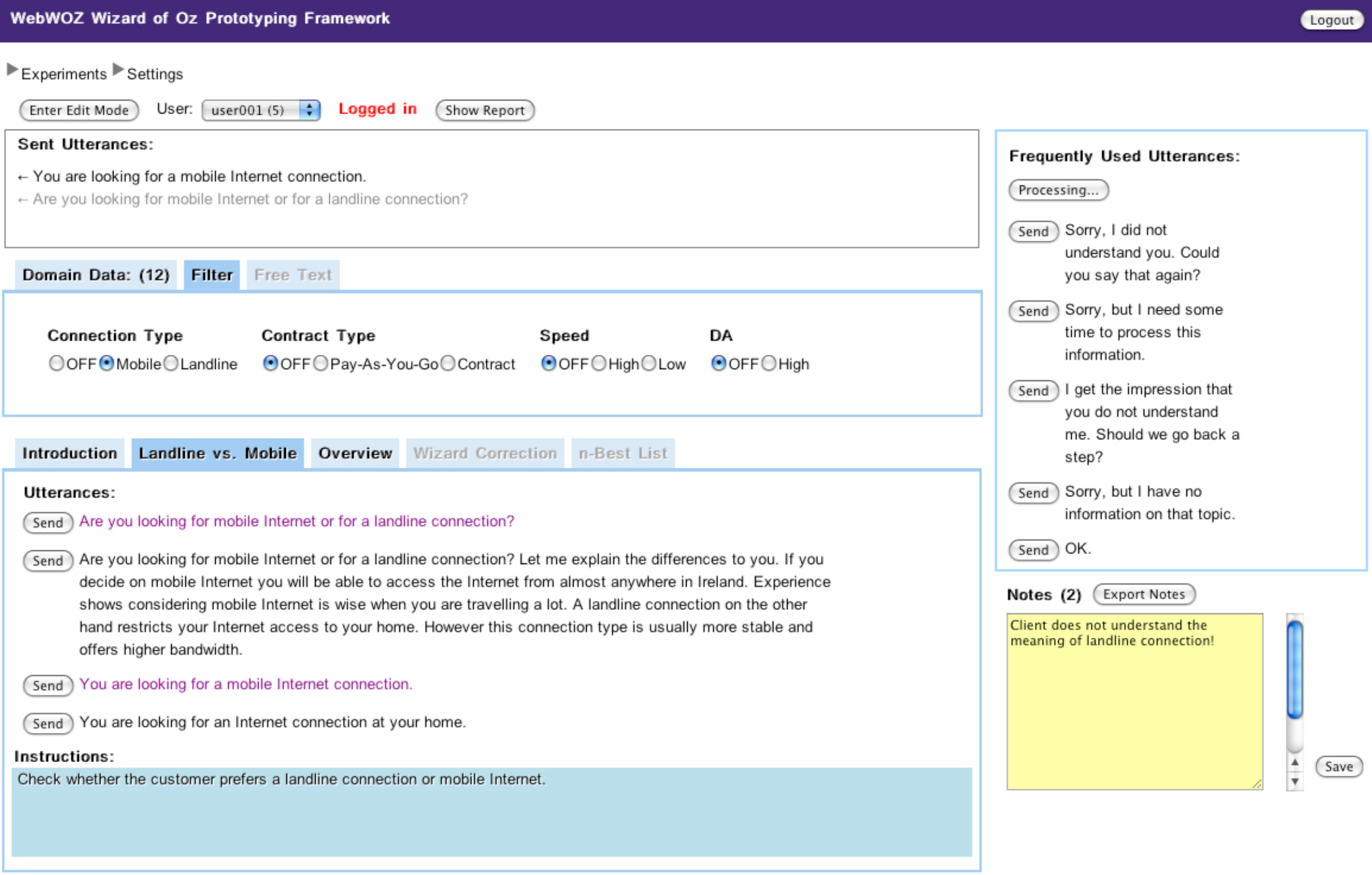} 
\caption{A wizard interface based on the generic framework.}
\label{fig:webwozwizard}
\end{figure*}

Similar to the wizard interface used with our initial prototype, this new platform is based on a staged dialogue structure (Fig.~\ref{fig:webwozwizard}). However, several additional features were integrated in order to address some of the problems identified. First, the dialogue structure was implemented using a tab layout, to make it easier to distinguish the different dialogue stages and track dialogue progress. Second, the area for recovery utterances was converted into a more general place-holder for frequently used utterances. Third, editing functions were implemented to allow the wizard to add, edit and delete utterances as well as move them between different dialogue stages or mark them as frequently used. Furthermore, in order to provide more freedom when interacting with a test participant, the new interface offers the possibility of including a chat-style text input field. Whereas in most WOZ experiments this kind of free interaction should be avoided, in some situations the experimenter may wish to explore the design space more freely. 

With a fourth feature, we tried to tackle the problem of retrieving domain specific information. A configurable filter mechanism for domain-data-based utterances was implemented that enables wizards to specify filters as well as filter values themselves, giving them better control over the domain data. Furthermore, a history element was introduced which holds all of the utterances sent to a test participant. Utterances are listed in chronological order and preceded by an arrow to the left, marking them as outgoing; the most recent sentence being highlighted in a different colour. In cases where the ASR component is used, its output is also displayed in the history, preceded by an arrow to the right, marking it as incoming. Finally, in order to alleviate timing problems, a notification system was implemented to inform the wizard that a test participant is ready to begin. Areas for taking time-stamped notes and fields for reminders and instructions were also added, with the aim of facilitating experimental analysis and improving the consistency of wizard behaviour.

Our new platform also allows a wizard to choose from a set of pre-configured language components to be used in an experiment. It is possible to choose which components to turn on and which to turn off. Additional component settings that are currently supported include the language for ASR and TTS as well as the language pair, i.e. the source and the target language, for an MT component. If an experiment setting requires a more controlled set-up in which MT or TTS (or both) need to be consistent, pre-defined translations and pre-recorded audio can be added. 

In cases where the wizard is to play an augmentation role, they can choose whether component output is immediately sent to a test participant, or whether it is post-edited first (supporting the integration of unreliable components). Two modes of augmentation are supported. In N-best list mode the wizard can choose from a list of possible outputs. This mode is particularly useful in situations where a component does produce understandable results, but the output is ambiguous. In cases where the component quality is considerably flawed, we also provide a correction mode in which a wizard is able to fully edit the output before it is sent. Both modes are supported for ASR as well as MT components. 

Having built a more generic WOZ prototyping platform for language technology applications our next goal was to test whether potential wizards were able to design experiments and if so, whether they would make use of some of the features we had integrated with this tool. The next sections of this paper therefore report on two rounds of evaluations where experts and non-experts were asked to use the platform to implement two different WOZ experiments.

% =============================================================================
\section{Supporting Experiment Construction}
% =============================================================================
Using this generic WOZ prototyping platform a new set of evaluations was conducted. The goal was to explore the feasibility of constructing experiments. Having a low overhead on experiment construction was earlier identified  as an important requirement (cf. Section~\ref{sec:requ-wizard-oz}). Therefore our WOZ prototyping platform introduced the possibility to quickly create and configure a new experiment as well as add, edit and delete dialogue utterances. Unrestricted interaction (if desired) is supported by the use of a chat-box, and generic filtering allows for quickly browsing different sorts of domain data. The purpose of integrating these features was to allow wizards to design WOZ experiments themselves without the need for any programming experience; comparable to a website Content Management System (CMS). 

% =============================================================================
\subsection{Evaluation Method}
% =============================================================================
The first stage of evaluation was a proof of concept activity that looked at the coverage of the given design space. The goal was to test whether each of the plausible configurations identified earlier (cf. Table~\ref{table:designspace}) could be supported without adapting the platform. Instances integrating various different component set-ups could be realized, achieving our goal for comprehensive WOZ tool support. While it was not our goal to actually run experiments with all of these instances (i.e. we were mainly interested in whether we could build them), it provided validation of the range of possible use cases the prototyping platform would be able to support. The published literature was used as a source for realistic product ideas. Elaborating on these ideas we built a total of 8 different WOZ experiments, all of which were using a combination of at least two different language technologies (i.e. ASR, MT, TTS), along with variants of these to cover all plausible language technology component configurations. Both the experiment structures as well as the relevant text utterances were created.

In a next step we wanted to test whether also other people (i.e. potential wizards) would be able to design experiments. To do so we conducted two sets of user studies. First we recruited researchers from the NLP area (expert users) and asked them to design two different WOZ experiments. Second, in order to test whether our tool would also enable non-experts to create WOZ experiments, we ran another study with participants from outside the NLP domain. The following sections report on the results of these two evaluations.

% =============================================================================
\subsection{Construction Study 1 - Use by Expert Users}
% =============================================================================
In order to validate the experiment creation process of our WOZ platform we conducted a study with 10 researchers working with language technologies. None of them was familiar with our tool; five of them had experience with WOZ. They were given a short introduction to the prototyping platform and a written manual they could refer to. Following registration with the platform they were asked to carry out two design tasks. For the first task they were given the exact wording of 16 different response utterances for a potential phone banking application. They had to add these utterances to a new experiment, arrange them in a useful way, and add any utterances they thought might be missing. The second task was to design a pizza ordering system. This time participants were not given any utterances and therefore had to come up with their own designs. The tasks were intended to be progressive - the first was relatively closed, the second open ended and allowing for more creativity. Descriptive statistics are provided for each task, but are not intended for comparison purposes.
%The overall goal of this evaluation was to see whether participants would be able to add, edit and arrange utterances without having first participated in a real training session. Furthermore we were interested in whether they would make use of two genuine features of our platform, namely the tabbed dialogue structure and the dedicated area for frequently used utterances.

% =============================================================================
\subsubsection*{Success Rate and Complexity of Designs}
% =============================================================================
Overall participants did not experience any particular difficulties in achieving the given tasks. All were able to complete the first task within the given time frame of 30 minutes and could present their design for the second another 30 minutes later.
Looking at the produced dialogues in more detail we could see that for the first task two participants did not create all of the 16 utterances they were given (i.e. P02 missed one utterance and P08 left out four) and one participant added to them (i.e. P07 created 22 instead of 16 utterances). The additions were caused by three given utterances that were each split into two separate ones, one additional question, one additional confirmation utterance and one additional advice utterance. 

For the second task we could see that participants used the previous example as a guideline. 
%On average they produced roughly the same number of utterances (mean=16.20, median=15) and an average of 11.51 (median=10.50) words per utterance in Task 1 compared to 10.47 (median=10.00) words per utterance in Task 1.
They produced on average 16.20 (SD=6.97) % median=15,
utterances where each of the utterances consisted of on average 11.51 (SD=3.64)%median=11.57, 
 words. The solutions produced were reasonable in the sense of allowing the task to be completed. A more complex task would likely result in more utterances.
\MNOTE{This sentence is confusing. Is the mean number of utterances for task 1 the same as for task 2? 'Roughly' doesn't sound very scientific; I think you should give the actual figures. I'm not sure the medians are very informative in this case; they could probably be omitted.}

This suggests that the produced dialogues exhibit approximately the same level of complexity. However, as one would expect from the more individualistic task setting, participants produced more varied designs in Task 2. 
In Task 1, the standard deviation (SD) of the number of utterances was 2.60 (mean=17.10), %, median=17
 while in Task 2 SD was 6.97 (mean=16.20). %, median=15
As regards number of words per utterance, for Task 1 we have SD=1.12 (mean=10.47), % , median=10.80
and for Task 2: SD=3.64 (mean=11.51).%median=11.57
\MNOTE{SL: reworded; please check.} 
As they were allowed to create their own designs, some participants showed better performance and more creativity than others. Overall, however, the results of this study show that expert participants were capable of handling both tasks successfully without additional upfront training.

% =============================================================================
\subsubsection*{Task and Usability Feedback}\label{sec11.2.2}
% =============================================================================
In order to obtain additional feedback with respect to the task we used post-task questionnaires to measure task difficulty as well as task satisfaction (cf. Table~\ref{table:questionnaires}). For the first a single 7 point Likert item running from 1 \textit{very difficult} to 7 \textit{very easy} was employed. The results show that the tasks were perceived as easy to complete leading to means of 6.2 for both the first %mode=6,
%median=6,
(SD=0.4) as well as for the second task %mode=7, median=6.5,
(SD=0.9). To measure task satisfaction we used three questions (i.e. 1. I am satisfied with ease of completion, 2. I am satisfied with the amount of time it took, and 3. I am satisfied with the supporting information) and again employed a 7 point Likert scale, this time running from 1 \textit{strongly agree} to 7 \textit{strongly disagree}. Also here the feedback was overall positive with means of 2.3 (SD 0.8), 2.5 (SD 1.0) and 2.3 (SD 1.5) for the three questions for the first task and 1.8(SD 0.4), 2.1 (SD 0.9) and 2.1 (SD 1.1) for the second task.

%(mode=3\textbar 2\textbar 1, median=2.5\textbar 2\textbar 2.5, SD=0.8\textbar 1\textbar 1.5) for the first and 1.8\textbar 2.1\textbar 2.1 (mode=2\textbar 2\textbar 1, median=2\textbar 2\textbar 2, SD=0.4\textbar 0.9\textbar 1.1) for the second task. 

\begin{table*}
\processtable{Questions used to evaluate the construction of WOZ experiments by expert users and learners.\label{table:questionnaires}}
{\begin{tabular*}{\textwidth}
{l | l}
\toprule
\multirow{3}{5cm}{Post-task questionnaire difficulty [7 point Likert item running from 1 \textit{very difficult} to 7 \textit{very easy}]} 
& \\
& Overall, this task was?\\
& \\
\midrule
\multirow{5}{5.1cm}{Post-task questionnaire satisfaction [7 point Likert item running from 1 \textit{strongly agree} to 7 \textit{strongly disagree}]} 
& \\
& Overall, I am satisfied with the ease of completing this task.\\
& Overall, I am satisfied with the amount of time it took to complete this task.\\
& Overall, I am satisfied with the provided support information when completing this task.\\
& \\
\midrule
\multirow{10}{5cm}{Post-test questionnaire usability (i.e. SUS) [7 point Likert item running from 1 \textit{strongly agree} to 7 \textit{strongly disagree}]} 
& I think that I would like to use this system frequently.\\
& I found the system unnecessarily complex.\\
& I thought the system was easy to use.\\
& I think that I would need the support of a technical person to be able to use this system.\\
& I found the various functions in this system were well integrated.\\
& I thought there was too much inconsistency in this system.\\
& I would imagine that most people would learn to use this system very quickly.\\
& I found the system very cumbersome to use.\\
& I felt very confident using the system.\\
& I needed to learn a lot of things before I could get going with this system.\\
\midrule
\multirow{3}{5cm}{Additional open questions} 
& These are problems I had using WebWOZ:\\
& These are features I would like to have added to WebWOZ:\\
& Here are some final recommendations:\\
\botrule
\end{tabular*}}{}
\end{table*}
 
In addition we asked participants about their interactions with the system and what types of features were missing. Looking at these results we see that at least some of them (2 out of 10) would have liked a more direct way to connect utterances and hence enforce a sequential order. Most of them, however, liked the freedom of independent utterances. Additional features that were recommended include drag-and-drop support to arrange utterances as well as instant feedback on when utterances are successfully stored in the system. Finally, an import function to add utterances from third party applications and more intuitive labels to reduce ambiguity were requested. Overall, however, the results of the evaluation indicate that participants liked the platform and that they were able to use it easily without having received extensive preceding training. 

In order to compare these results to an industry standard we calculated the respective System Usability Scale (SUS) score. This was done by first normalizing the scales of our SUS questionnaire i.e. points from positive statements (questions 1, 3, 5, 7 and 9) were reduced by 1 and points from negative statements (questions 2, 4, 6, 8 and 10) were subtracted from 7. Next, the resulting numbers were summed up and multiplied by 5/3 (Note: this converts a 0-60 points scale to a 0-100 points scale). Taking the average score across our 10 experts this led to a value of 77/100 (95\% confidence interval ranging from 68.75 to 85.25), corresponding to ``good''  on an adjectival scale~\citep{Bangor09}. For further information on how to compute SUS scores the reader is referred to~\cite{Bro96}.

%Calculating the System Usability Scale (SUS) score by aligning the scales (i.e. points from positive statements were reduced by 1 and points from negative statements were subtracted from 7) and normalizing them (the sum of all points was multiplied by 5/3) to convert them to values from 0 to 100 confirmed these results.

%The average System Usability Scale (SUS) score (i.e. the mean score computed across the 10 expert users cf.~\cite{Bro96}) of 77/100 (95\% confidence interval ranging from 68.75 to 85.25) produced by the post-test usability questionnaire seems to confirm these results, corresponding to ``good''  on an adjectival scale~\citep{Bangor09}.

% =============================================================================
\subsubsection*{Use of Features}
% =============================================================================
In addition to obtaining feedback about the usability of the platform we were also interested in whether participants would use the range of features available. One aspect here was the possibility to create a structured dialogue, i.e. using tabs to organize a dialogue in different stages. Looking at the log-files we found rather strong differences in people's preferences on that matter. While on average a participant created 3.71 (median=4) tabs per task the results highlight a spread going from 1 tab only to creating 8 (SD=2.63). However, it seemed that people structured the dialogue more in the second task (median=3 for the first task vs. median=4 for the second task) when they had to produce their own selection of utterances, i.e. when they had more freedom to design the interaction.
%As participants were not asked to actually run any experiments with their created dialogues, we are unable to tell whether they would have performed some fine-tuning after conducting a number of trial-runs, however we believe over time the number of tabs would have harmonized.\\

%s <- read.csv('exp_featureuse.csv')
%dev.new(width=8, height=5)
%plot.new()
%hist(s$TABS, xlab="", ylab="Absolute requency", main="Number of tabs")
%dev.copy2pdf(file='fig/exp_tabs.pdf', paper='special', family='Helvetica')
%
%s <- read.csv('exp_featureuse.csv')
%dev.new(width=8, height=5)
%plot.new()
%hist(s$FREQ, xlab="", ylab="Absolute frequency", main="Number of frequently used utterances", breaks=30, xlim=c(1,30))
%dev.copy2pdf(file='fig/exp_frequtt.pdf', paper='special', family='Helvetica')

\begin{figure*}
\centering
\includegraphics[width=0.7\textwidth]{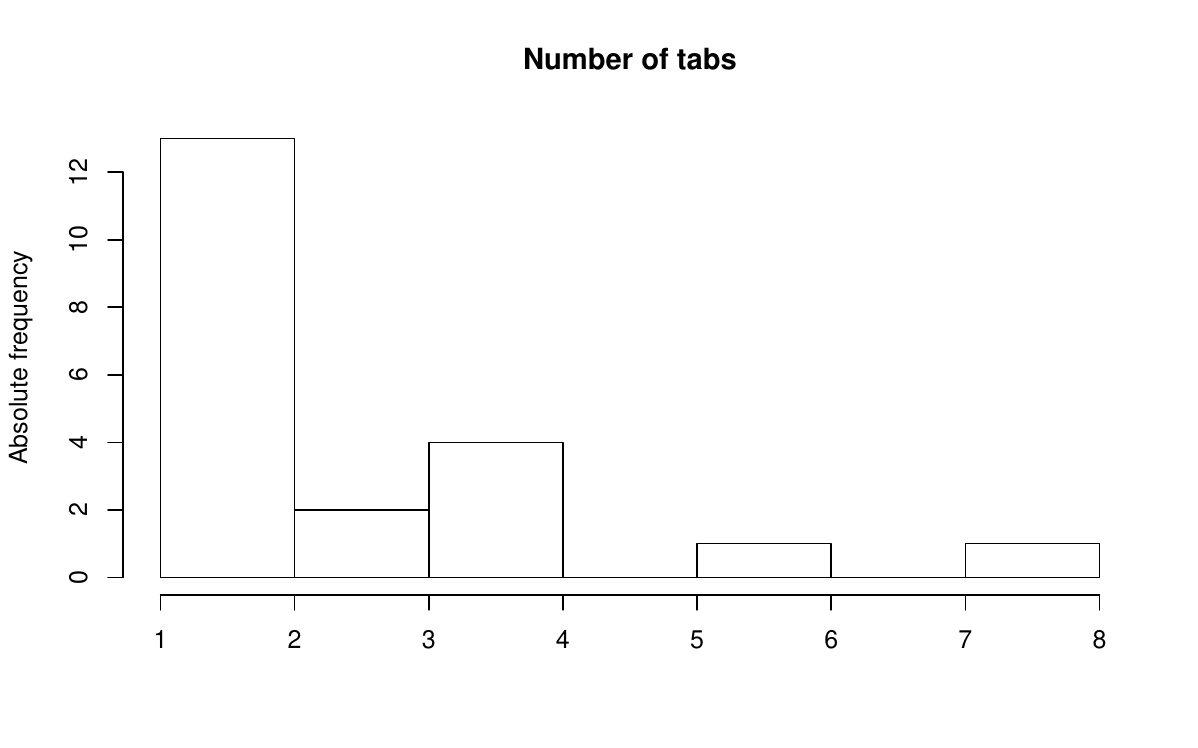} 
\includegraphics[width=0.7\textwidth]{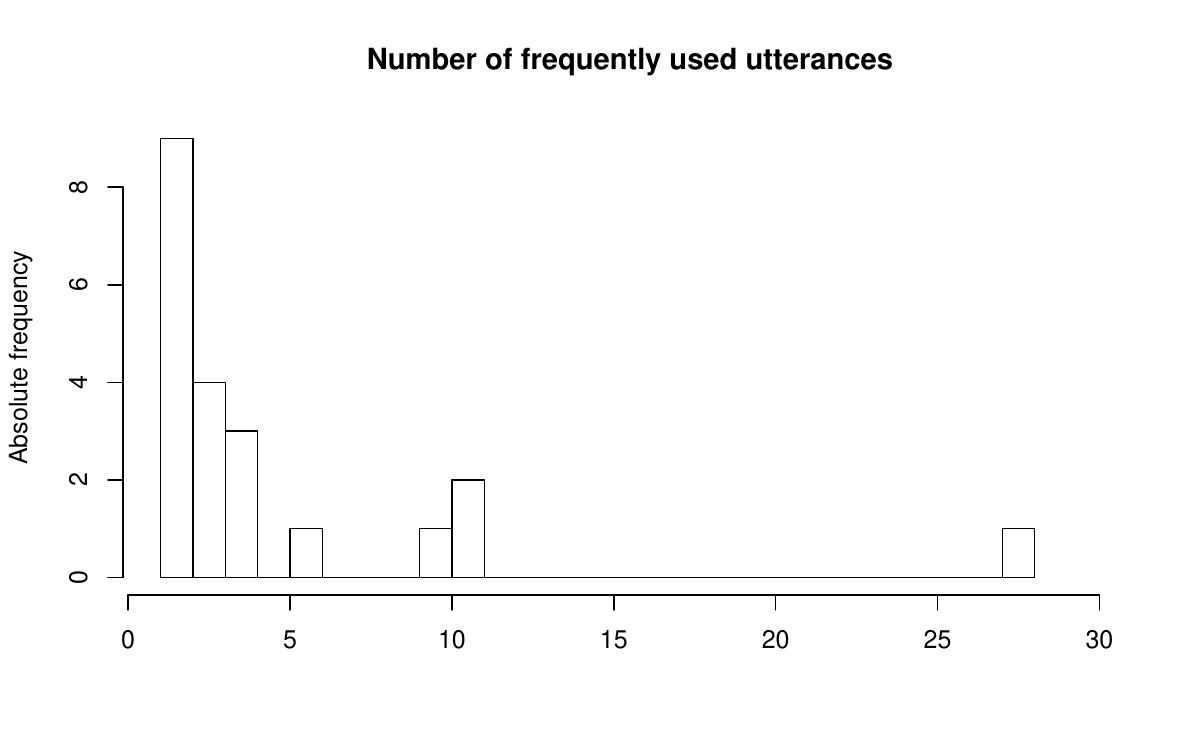}
\caption{Use of tabs and frequently used utterances by expert users.}
\label{fig:exp_features}
\end{figure*}
A second aspect we were interested in was the use of frequently used utterances. The use of this feature shows whether participants actually differentiated between utterances that can be dedicated to a specific dialogue stage and those that could be used several times throughout an interaction. Also here the log-file revealed a rather varied behaviour. On average participants defined 4.86 utterances as frequently used (median=3, SD=6.21) but with an overall range of up to 28 frequently used utterances. When asked about their understanding of the concept participants were in favour of the idea. However, we believe that without actually running experiments it was difficult for them to identify those utterances that are unlikely to be associated with only a single dialogue stage, which probably influenced the use of this feature. Furthermore, it needs to be highlighted that the limited amount of time (i.e. 30 minutes for designing a complete dialogue) most likely reduced the total number of utterances created. A more intense engagement with the design process might have resulted in more complex solutions, and consequently led to the inclusion of utterances that may be less obvious. Figure~\ref{fig:exp_features} shows the experts' use of tabs and frequently used utterances. 

In summary this first round of exploring the creation process showed that potential wizard users were able to successfully design experiments with our WOZ platform. They used different dialogue steps and understood the concept of frequently used utterances. 
%We believe that actually running experiments would probably have led to various adaptations both in the structure of the dialogues as well as the number of utterances used. However, a test scenario where both the design as well as the conduct of an experiment was evaluated, would have required significantly more time from our participants. In addition, as in such an experiment design they would have acted as wizards, we would have needed to provide them with (`dummy') test participants, which seemed not only due to the number of necessary people but also due to the lack of control we would have had over their actions, unrealistic to achieve. Furthermore, design adaptation and an increased usage of tabs and frequently used utterances might only be observed over time, which would have required an even greater number of test runs. Such an exhaustive evaluation was unfortunately out of the scope of this research agenda and hence we decided to keep design phase and experiment phase separated and rather increase the diversity of users that might be confronted with WOZ experimentation. Therefore, while for this initial evaluation of the creation process we had mainly recruited potential users who were familiar with the WOZ method, we conducted a second round of tests for which the goal was to expand the potential user base into other areas. 

% =============================================================================
\subsection{Construction Study 2 - Use by Learners}
% =============================================================================
While at least some researchers in NLP and HCI might be familiar with WOZ prototyping (cf. Section~\ref{sec:users}), people working in other fields are rarely exposed to the method. However, as described earlier, various application areas can benefit from human simulation. In particular, as language technologies are increasingly used in a variety of devices in combination with other input and output modalities, distinct contributions from different disciplines are required to offer intuitive and novel design solutions. Hence, one goal of the platform is to reduce this entry barrier and make WOZ prototyping accessible to people outside the field of NLP research.

%%R code for chart
%
%s <- read.csv('exp_nonexp_wc.csv')
%dev.new(width=8, height=5)
%plot.new()
%boxplot(s$wc ~ s$type, xlab="User", ylab="Average word count per utterance", col="bisque", range=0)
%dev.copy2pdf(file='../fig/exp_nonexp_wc.pdf', paper='special', family='Helvetica')
%

%\begin{figure*}
%\centering
%\includegraphics[width=0.7\textwidth]{fig/exp_nonexp_wc.pdf} 
%\caption{Comparing experts and non-experts on the complexity (average words per utterance) of created utterances.}
%\label{fig:expnonexp_wc}
%\end{figure*}

In order to test whether our approach of integrating CMS-like features into a WOZ platform makes it easy enough to be used by novices, we conducted tests with 51 student participants. Students' backgrounds included Computer Science, Information Systems as well as HCI. While the tests were conducted during class, participation was entirely voluntary and did not involve any actual course credit or other type of compensation. As HCI teaching is one of the purposes envisaged by the tool, this sample is seen as a reasonable starting point. However it would be desirable to consider other learner demographics in future work.
%No NLP people were recruited this time.
Participants were asked to perform the same two tasks as their predecessors.
% and in addition asked to complete a feedback form.

% =============================================================================
\subsubsection*{Success Rate and Complexity of Results}
% =============================================================================
12 of the 51 participants were excluded as they either did not give us the necessary consent to use their data or did not submit any designs. From the remaining 39 participants, 28 worked on both design tasks.

\begin{figure*}
\centering
\includegraphics[width=0.7\linewidth]{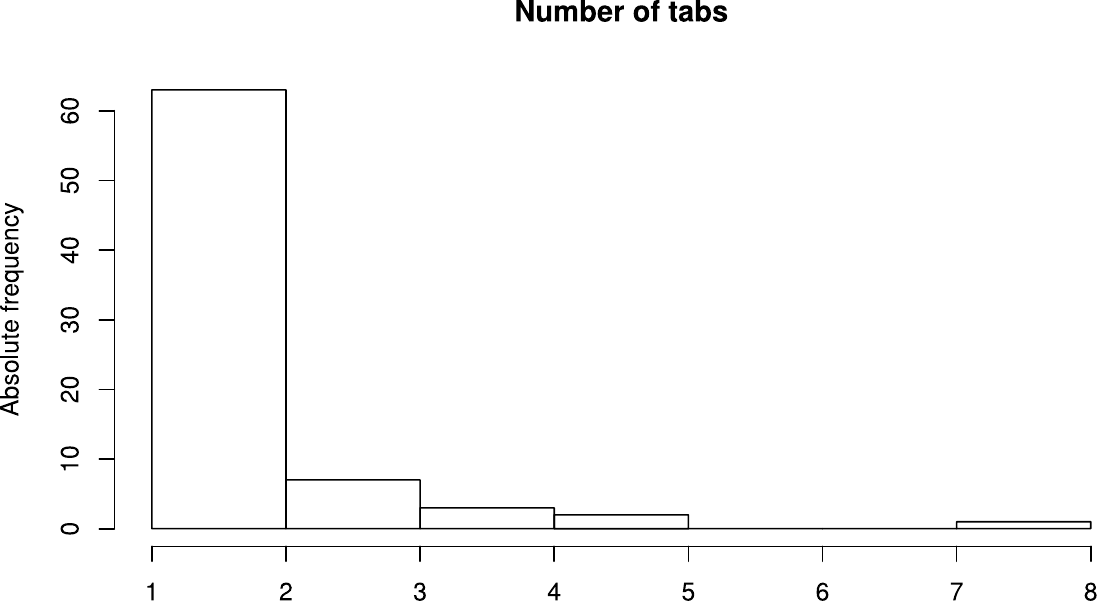}\includegraphics[width=0.7\linewidth]{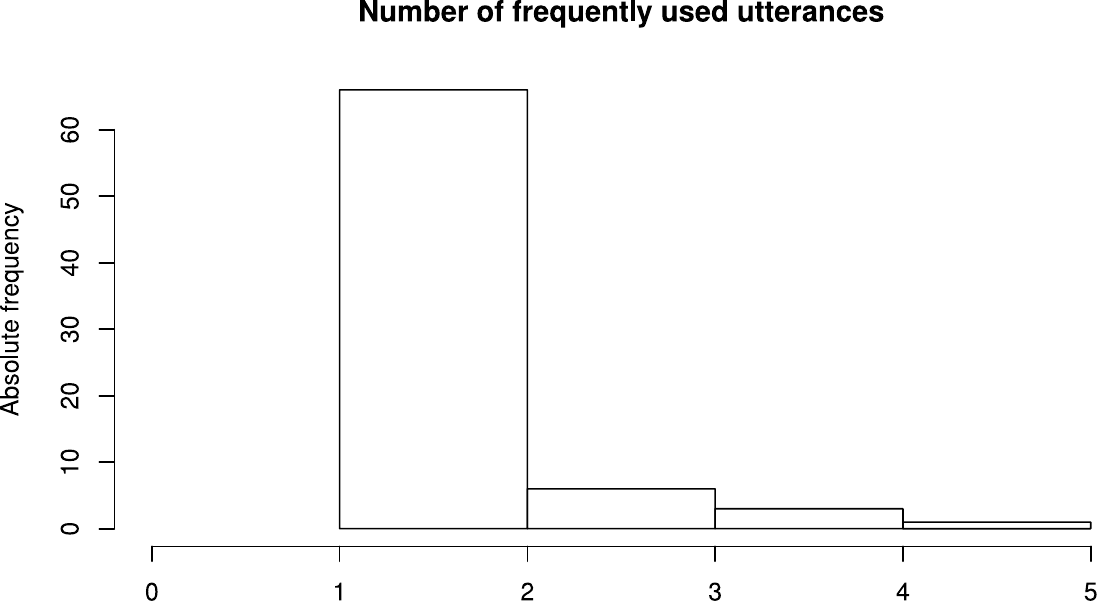}
\caption{Use of tabs and frequently used utterances by learners.}
\label{fig:nonexp_features}
\end{figure*}

%From the remaining 39 participants, 28 were able to finish the first task and work on the second within the given time-frame.
%If we compare the results to what the `expert' users had achieved in the previous study, we see that in terms of the complexity of the produced designs the two groups of users are fairly close in their performance.
In the first task our 39 students produced on average 12.36 utterances (median=15.00, SD=6.47) with an average words/utterance rate of 10.36 (median=10.80, SD=2.78). With respect to completeness, 14 participants created all the utterances requested in task 1, 21 created fewer utterances, 
%(average=7.91, median=7.00, SD=4.96)
and 3 created more.
% (average=22.00, median=21.00, SD=2.65). 
The 28 students who moved on to the second task additionally created on average 14.39 utterances (median=12.50, SD=7.93) with an average words/utterance rate of 10.85 (median=10.33, SD=3.42). While the sample sizes preclude any comparative analysis, these figures appear broadly in line with those of expert users. Also the use of tabs was approximately the same, so that only the number of frequently used utterances suggests a clear difference between expert users and learners (cf. Figure~\ref{fig:nonexp_features}).

%n <- read.csv('nonexp_featureuse.csv')
%dev.new(width=8, height=5)
%plot.new()
%hist(n$TABS, xlab="", ylab="Absolute frequency", main="Number of tabs")
%dev.copy2pdf(file='fig/nonexp_tabs.pdf', paper='special', family='Helvetica')
%
%n <- read.csv('nonexp_featureuse.csv')
%dev.new(width=8, height=5)
%plot.new()
%hist(n$FREQ, xlab="", ylab="Absolute frequency", main="Number of frequently used utterances", breaks=5, xlim=c(0,5))
%dev.copy2pdf(file='fig/nonexp_frequtt.pdf', paper='special', family='Helvetica')

\subsubsection*{Task and Usability Feedback}
% =============================================================================
Looking at the subjective task and usability feedback we received from the student participants via post-task and post-test questionnaires (cf. Section~\ref{sec11.2.2} ) we see an increase in perceived task difficulty for the first (mean=4.3, 
%mode=6, median=4,
SD=1.7) as well as for the second task (mean=4.8,
%mode=6, median=5,
SD=1.6). Also, people were less satisfied with ease of completion (Task 1: mean=3.5,
%mode=3, median=3,
SD=1.7; Task 2: mean=2.8,
%mode=2, median=2,
SD=1.6), the time it took them to do so (Task 1: mean=3.1,
%mode=2, median=2,
SD=1.7; Task 2: mean=3.1,
%mode=2, median=2.5,
SD=1.8) as well as the support information they were provided with (Task 1: mean=4,
%mode=3, median=4,
SD=1.9; Task 2: mean=3.6,
%mode=5, median=3,
SD=1.7).
The same result is reflected by the questionnaire assessing the overall tool usability which showed an average SUS score of 58/100 (95\% confidence interval ranging from 51.64 to 64.38) - substantially lower than the expert group.
In summary the results of this evaluation show that while the design of WOZ experiments does require a certain amount of knowledge and
understanding of the method (with learners perceiving the tasks as more difficult to carry out with the tool), even without dedicated training students were able to produce workable solutions. Easy access to relevant tools such as our WOZ platform may therefore be seen as an enabler which can open up the design space of WOZ prototyping to people outside the field of language technologies.

% =============================================================================
\section{Supporting WOZ Experiments}
% =============================================================================
One goal of our prototyping platform is to reduce the effort needed to design and conduct a WOZ experiment. We have shown how both expert and novice users were able to design and build realistic experiments with the WebWOZ tool. In order to evaluate whether such a tool can also be used for real experimentation, two WOZ studies, both of which employed a single wizard interaction with various test subjects, were conducted. Each study provides an initial point of validation for the approach, corresponding to a particular part of the design space. 
% =============================================================================
 \subsection{Experiment Support Study 1 - Translated Speech Synthesis}
% =============================================================================
The first study built upon the results provided by our initial prototype study (cf. Sections~\ref{sec:tool}) and aimed at extending them into the spoken language domain. That is, this time it employed German text as well as synthesised speech output.
%which also permitted us to compare those two modes with respect to wizard consistency on sending utterances. 
The technical set-up was similar to the one described in Section~\ref{sec:set} using our WOZ platform alongside an active Skype\footnote{http://www.skype.com/ [Accessed: 23$^{rd}$ Dec. 2013]} session to transfer speech to the wizard from her test-participants. From an experimental point of view the person who acted as the wizard was interested in understanding the influence machine translated utterances have on synthesised speech output and how this changes the user experience when interacting with a system i.e. to what degree are flawed translations acceptable in an interactive speech-to-speech dialog setting. This is an instance of Case 9 of our analysis of the WOZ design space for Language Technology applications (Speech input, input MT, DM, output MT, speech output, cf. Table	~\ref{table:designspace}). Our German speaking wizard replaced the ASR and input MT and then chose an appropriate response from a list of possible response utterances.

For this very specific experiment setting it was also important to control quality for both output MT and TTS. Hence, possible response utterances were pre-translated from English into German using two different online translation systems (i.e. Google Translate\footnote{http://translate.google.com/ [Accessed: 23$^{rd}$ Dec. 2013]} vs. Systran\footnote{http://www.systranet.com/translate/ [Accessed: 23$^{rd}$ Dec. 2013]}), synthesised and stored in the system. The wizard was then restricted to choosing these utterances based on the dialog progress (Note: in the wizard interface the utterances were still displayed in English, on the client side, however, the corresponding German utterances were displayed and spoken out). Unrestricted, free-form text was not available for the wizard. The 17 test participants who took on the role of test customers were all native speakers of German. They were told that they would be interacting with a system that understands spoken input in German and asked to solve two information retrieval tasks similar to those used in the previous study (cf. Sections~\ref{sec:tool}). In one of the tasks the system would communicate with them via German speech output, in the second, it would produce German text on a screen.

After the experiment participants were informed that a human wizard operated the system. Additional details and results of this study exploring the influence MT has on TTS can be found in~\citep{Schneider13}. 

% =============================================================================
\subsection{Experiment Support Study 2 - Pronunciation Trainer}
% =============================================================================
The second study employed WOZ to collect a corpus of realistic dialogue utterances for an online language pronunciation trainer. For this purpose we were working with researchers from a different institution who had already developed a working prototype of a system that could analyse a test-participant's pronunciation of an English sentence and highlight which words or parts of a sentence were mispronounced. Linking this analysis to actual textual feedback was, however, not supported at the time. The study therefore used a human wizard to produce real-time textual feedback based on the results of the pronunciation analysis. Feedback was provided in English and aimed at pointing language learners at the words or parts of words that were mispronounced. All language learners were able to understand English. The envisioned tool should simply be used to improve their pronunciation and not to teach them new vocabulary.

With respect to the design space for WOZ in language technology applications such a setup closely resembles the simulation of a dialog system which accepts spoken input and provides text-based feedback (Case 2: speech input, no MT, text output, cf. Table~\ref{table:designspace}). Our WOZ tool was used to implement the study. Different text elements were prepared so that they could be assembled to flexibly form a feedback sentence. The wizard was able to compose the sentence and fill in the details or alternatively create a response completely from scratch. On the client side the pronunciation system was integrated with the WOZ client interface (i.e. it was running in a separated frame integrated in the web-based pronunciation training system). Again, Skype was used to transfer the spoken input from a test-participant to the wizard. Feedback sent from the wizard was then displayed in a text box situated in the bottom of the screen. A member of the other research team, who was also involved the development of the pronunciation analysis system, acted as a wizard. One trial run was conducted to test the set-up after which 12 test-participants were recruited to train their pronunciation. Additional details and results of this study  can be found in~\citep{Cab12}.

\vspace{5 mm}
\subsection{Remarks on the use of WebWOZ in live experiments}
\MNOTE{SL: made this a separate section to emphasize 'lessons learned' from these two studies (from the WebWOZ perspective, that is). Maybe the text below could be further enhanced.}%
\MNOTE{SS: I added the reference to the new EICS paper.}
These two studies provide initial validation of the suitability of the WebWOZ prototyping platform for real experimentation, and of this overall approach to supporting WOZ. They also provide evidence that the built-in customization mechanisms (i.e. integration of CMS-like features for the selection of specific language technology settings), and adaptability (i.e. its use of web technologies to permit integration with existing experiment environments), allow for an employment in a variety of settings. Additional work that aims at further improving the broad application of the platform for language technology research is already in progress. To do so the system has  been installed in other environments where it will serve as a tool in several research projects \citep[e.g.][]{Sch13, Mil13}. 

\section{Future work}
Future work will examine in greater depth the consistency and performance aspects of the wizard's task and how these issues can be addressed. Our interviews with wizards running experiments showed that they are aware of these aspects but struggle to control them. 
Investigating these aspects will require an extensive program of experimentation with multiple wizards running their own experiments, and as part of this we plan to make the system available to the wider HCI community, for both teaching and research purposes. As a first step the current version of the platform has been published under the Apache License (Version 2.0) and is available for download\footnote{https://github.com/stephanschloegl/WebWOZ, http://www.webwoz.com [Accessed: 23$^{rd}$ Dec. 2013]}. The small interview study conducted as part of the analysis yielded many interesting insights, and so a more substantial qualitative research study would be worth pursuing. 

In addition, steps towards the inclusion of multi-modal aspects have been undertaken by integrating video output. With some creativity, we believe, the platform could also be used to explore a number of richer multi-modal scenarios. Likewise, the advent of mobile web browsers opens up a number of interesting possibilities, particularly in the context of speech-to-speech translation~\citep{Stu06} - a domain where the platform, due to its web-based nature, may easily be deployed. It would also be interesting to run comparative experiments between WebWOZ and other available tools.

With respect to the exploration of the WOZ method, another interesting direction is the provision of feedback to wizards on their own consistency, along with metrics such as response time and spread of utterances used. This might also play a role in training and piloting, allowing the wizard to decide at what point she has practised the dialogue enough and is ready to commence the full experiment. Also, experiments with multiple wizards represent an important methodological aspect (one that was already highlighted numerous times in the literature) that needs further exploration. Finally, from a language technology point of view different aspects of dialogue management (DM) should be investigated. So far we have treated DM as an integrated wizard task including language understanding as well as output generation. However, these are distinct components and research topics which are worthy of more focussed consideration.

% =============================================================================
\section{Conclusion}
% =============================================================================
This paper has examined the Wizard of Oz method in general and its use for simulating natural language-based interaction in particular. Drawing on a survey of the literature and a focused interview study with people who have used the method, it was highlighted that existing tools provide insufficient support for this form of prototyping and that both a better understanding of the method and more flexible tools are needed. 

Following a systematic analysis of the design space and its influence on the wizard's task we presented several requirements which seem crucial to improve future tool support. The WebWOZ prototyping platform was presented which integrates different features for dealing with those requirements. First, to reduce the overhead in experiment construction and software installation as well as to increase the flexibility in terms of supporting both highly structured and more exploratory experimentation, a web-based CMS-like wizard interface was implemented. Second, a flexible integration of language technologies via web-services was demonstrated and it was further outlined how this architecture would support a possible augmentation role of the wizard. %Third, several data logging and export mechanisms were integrated supporting the analysis of experiment results. 

Evaluations suggest that the approach is workable. We showed that potential users can construct WOZ experiments quickly, which is critical if a technique comparable to `sketching' \citep[cf.][]{bib:Buxton07s} is to be supported, and that a comprehensive coverage of the design space can be achieved; supporting the complete list of scenarios outlined in Table~\ref{table:designspace}. It should be noted that these scenarios also cover potential multilingual experiments. Thus, machine translation was added as a dedicated component to the natural language interaction pipeline, setting our WebWOZ prototyping platform apart from most currently existing WOZ systems. 

Focusing more generally on the challenges of WOZ, our evaluations suggest that wizards require additional support. One difficulty concerns the selection of appropriate response utterances where, despite sufficient familiarisation with a simulated dialogue, a wizard might face the challenges of consistency and timing. A second source of potential error arises through insufficient awareness, which was also identified as a cause for inconsistent wizard actions.

\ack
This research was supported by the Science Foundation Ireland (Grant 07/CE/I1142) as part of the Centre for Next Generation Localisation (www.cngl.ie) at Trinity College Dublin.

\makeatletter
\def\@biblabel#1{}
\def\newblock{}
\makeatother

%\bibliography{iwcomp}
\bibliographystyle{apalike}

\end{document}